\documentclass[a4paper,11pt,hyper]{JHEP3}
\usepackage{amsmath,latexsym,amssymb}
\usepackage{graphicx}
\usepackage[latin1]{inputenc}
\usepackage{bbm}
\usepackage[stable]{footmisc}

\newcommand{\be}{\begin{equation}}
\newcommand{\ee}{\end{equation}}
\newcommand{\ba}{\begin{eqnarray}}
\newcommand{\ea}{\end{eqnarray}}
\renewcommand{\a}{\alpha}
\newcommand{\al}{\alpha}
\renewcommand{\b}{\beta}
\renewcommand{\d}{\delta}
\newcommand{\e}{\epsilon}

\renewcommand{\k}{\kappa}

\newcommand{\m}{\mu}

\newcommand{\Om}{\Omega}
\newcommand{\om}{\omega}

\newcommand{\s}{\sigma}

\newcommand{\p}{\partial}
\newcommand{\w}{\wedge}

\newcommand{\nn}{\nonumber}
\newcommand{\non}{\nonumber}
\newcommand{\hlf}{\frac12}
\newcommand{\lp}{\left(}
\newcommand{\rp}{\right)}
\newcommand{\ls}{\left[}
\newcommand{\rs}{\right]}

\newcommand{\eq}[1]{\begin{equation}#1\end{equation}}
\newcommand{\spl}[1]{\begin{split}#1\end{split}}

\def\dd{\text{d}}

\title{Moduli Stabilization and Cosmology of Type IIB on SU(2)-Structure Orientifolds}

\author{Claudio Caviezel${}^{\diamondsuit}$, Timm Wrase$^{\clubsuit}$ and Marco Zagermann${}^{\clubsuit}$\\
\begin{itemize}
  
\item  Max-Planck-Institut f\"ur Physik\\
       F\"ohringer Ring 6, 80805 M\"unchen, Germany
  
\item  Institut f\"{u}r Theoretische Physik \\
       \& Center for Quantum Engineering and Spacetime Research\\
       Leibniz Universit\"{a}t Hannover \\
       Appelstra{\ss}e 2, 30167 Hannover, Germany
\end{itemize}

\bigskip
Email: \email{caviezel@mppmu.mpg.de,\\ timm.wrase, marco.zagermann @itp.uni-hannover.de}}

\abstract{We consider type IIB flux compactifications on six-dimensional SU(2)-structure manifolds with O5- and O7-planes. These six-dimensional spaces allow not only for $F_3$ and $H_3$ fluxes but also for $F_1$ and $F_5$ fluxes. We derive the four-dimensional $\mathcal{N}=1$ scalar potential for such compactifications and present one explicit example of a fully stabilized AdS vacuum with large volume and small string coupling. We then discuss cosmological aspects of these compactifications and derive several no-go theorems that forbid dS vacua and slow-roll inflation under certain conditions. We also study concrete examples of cosets and twisted tori and find that our no-go theorems forbid dS vacua and slow-roll inflation in all but one of them. For the latter we find a dS critical point with $\epsilon$ numerically zero. However, the point has two tachyons and eta-parameter $\eta \approx -3.1$.}


\preprint{ITP-UH-21/09\\ MPP-2009-212}

\begin{document}

\setcounter{footnote}{0}
\renewcommand{\thefootnote}{\arabic{footnote}}
\setcounter{section}{0}

\section{Introduction}
Flux compactifications of type II string theory (see e.g. \cite{Grana:2005jc,Douglas:2006es,Blumenhagen:2006ci,Denef:2007pq} for reviews) have been intensively studied in the last couple of years. The simplest compactification spaces that lead to a four-dimensional $\mathcal{N}=1$ action with a non-vanishing scalar potential for the closed string moduli are Calabi-Yau orientifolds threaded with $H_3$ and RR fluxes. In type IIB such compactifications allow only for $H_3$ and $F_3$ fluxes, which generically fix all complex structure moduli and the dilaton and yield no-scale type Minkowski vacua at tree level with unfixed K\"{a}hler moduli \cite{Gukov:1999ya,Giddings:2001yu}. In order to fix also the K\"{a}hler moduli, typically in AdS-vacua, the no-scale property has to be broken, which may naturally happen due to non-perturbative quantum \cite{Kachru:2003aw} or perturbative $\alpha^{\prime}$ corrections \cite{Becker:2002nn,Balasubramanian:2004uy,Balasubramanian:2005zx}.

On the other hand, in type IIA string theory compactified on Calabi-Yau orientifolds threaded with $p$-form fluxes \cite{Louis:2002ny,Grimm:2004ua} it is possible to stabilize all geometric moduli at tree-level in AdS vacua \cite{DeWolfe:2005uu,Ihl:2006pp}. In \cite{Lust:2004ig} it was shown that supersymmetric AdS vacua in type IIA cannot only be obtained from Calabi-Yau compactifications but also from the more general class of SU(3)-structure manifolds. This has lead to the exploration of compactifications on SU(3)-structure manifolds \cite{Derendinger:2004jn,Villadoro:2005cu,House:2005yc,Grana:2005ny,Camara:2005dc,KashaniPoor:2006si,Grana:2006hr,Benmachiche:2006df,Koerber:2007xk,Ihl:2007ah,Cassani:2007pq,KashaniPoor:2007tr,Cassani:2008rb} for which it is also possible to stabilize all moduli at tree-level in AdS vacua.

Having compactifications with a tree-level scalar potential that depends on all closed string moduli, one may ask whether it is also possible to have meta-stable  de Sitter (dS) vacua or slow-roll inflation \cite{Hertzberg:2007ke}. In \cite{Hertzberg:2007wc} a no-go theorem was derived that forbids dS vacua and slow-roll inflation in type IIA Calabi-Yau compactifications with $p$-form fluxes and O6/D6-sources. A manifold with negative scalar curvature, however, could in principle evade this no-go theorem. Using this approach, four-dimensional dS vacua were constructed in \cite{Silverstein:2007ac,Haque:2008jz}, but it was   argued in \cite{Danielsson:2009ff} that it may be difficult to satisfy the underlying 10-dimensional equations of motion. A related approach was used in \cite{Caviezel:2008tf,Flauger:2008ad}, where the authors studied compactifications on coset spaces \cite{Nilsson:1984bj,Lust:1986ix,Castellani:1986rg,LopesCardoso:2002hd,House:2005yc,Aldazabal:2007sn,Tomasiello:2007eq,Koerber:2008rx,Chatzistavrakidis:2008ii} and twisted tori \cite{Hull:2004in,Dabholkar:2005ve,Grana:2006kf,Hull:2006qs,Hull:2006va,Cvetic:2007ju,Ihl:2007ah,Bergman:2007qq,Derendinger:2004jn,Villadoro:2005cu}. The authors derived several new no-go theorems that were used to exclude many concrete examples, but explicit dS extrema with one tachyonic direction were also found in both papers. Since this tachyonic direction is different from the one discussed in \cite{Covi:2008cn,Covi:2008ea} (see also \cite{Achucarro:2008fk}), its origin is unclear. In \cite{deCarlos:2009qm} an extensive search for dS vacua was performed in a related model and the authors found dS extrema. However, stable de Sitter (and Minkowski) vacua were only found in special non-geometric compactifications. So it remains an open problem to construct geometric flux compactifications that lead to controlled stable dS vacua at tree-level.

In this paper we investigate type IIB compactifications that classically lead to a four-dimensional $\mathcal{N}=1$ supergravity action that could possibly be interesting from a cosmological point of view. In particular, we would like to have a scalar potential that depends at least on all geometric moduli at tree-level and that potentially could allow for dS vacua or slow-roll inflation. This requires orientifold planes to evade the no-go theorems of \cite{Gibbons:1984kp,deWit:1986xg,Maldacena:2000mw} (see also \cite{Steinhardt:2008nk,Neupane:2009ws,Neupane:2009br}). We will take these O-planes to be smeared over their transverse directions (see \cite{Acharya:2006ne} for a discussion of this point). Furthermore, for simplicity we restrict ourselves to closed string moduli (i.e. we do not include D-branes), and only consider bulk moduli in the analysis of concrete examples. As we explain in section 5, this leads us to study flux compactifications on SU(2)-structure manifolds with O5- and O7-planes (see e.g.  \cite{Dall'Agata:2004dk,Behrndt:2004mj,Bovy:2005qq,Grana:2006kf,Kounnas:2007dd,Andriot:2008va,Lust:2009mb,Lust:2009zb,Triendl:2009ap,Louis:2009dq,Danckaert:2009hr} for flux compactifications on SU(2)-structure manifolds). In these models, also $F_1$ and $F_5$ flux can be turned on, as opposed to the warped Calabi-Yau compactifications of \cite{Giddings:2001yu}, where the absence of one- and five-cycles does not allow for these types of fluxes. As we show explicitly in one particular example, compactifications of this type can actually stabilize all moduli at tree level in a large volume and small string coupling regime.

As we discuss in more detail in section \ref{IIBcomp}, SU(2)-structure manifolds have two globally defined vector fields that lead to a natural (2+4)-split of the tangent bundle (in the special case of SU(2)-holonomy, even the whole manifold factorizes as $T^2\times K3$). Furthermore, the explicit examples we consider in this paper are all parallelizeable manifolds. This allows us to choose a basis of six globally defined vector fields, $e^m$ $(m=1,\ldots,6)$, that is compatible with the (2+4)-split (i.e., we take $e^1$ and $e^2$ to span the two-dimensional part of the tangent spaces singled out by the SU(2)-structure, and $e^3,\ldots,e^6$ in the four-dimensional complement). Aligning the orientifold planes with this adapted basis of vector fields, one finds that the most general O-plane setup that still preserves $\mathcal{N}=1$ supersymmetry is, up to permutations, given as in  table \ref{table:IIBOplanes}.

\begin{table}[h!]
\begin{center}
\begin{tabular}{|c||c|c||c|c|c|c|}
  \hline
  plane & 1 & 2 & 3 & 4 & 5 & 6 \\ \hline \hline
  O5 & & & x & x &  &\\ \hline
  O5 & & & & & x & x \\ \hline
  O7 & x & x & x & & x & \\ \hline
  O7 & x & x & & x & & x \\ \hline
\end{tabular}
\end{center}
\caption{O5- and O7-planes}\label{table:IIBOplanes}
\end{table}

This follows from the fact that for a pair of D-branes or O-planes to preserve a common supersymmetry, the number of mixed Neumann-Dirichlet boundary conditions must be divisible by four. Moreover, the globally defined vector fields of the SU(2)-structure must be even under the O7-projection and odd under the O5-projection, which explains the orientation of the O-planes in the (1,2)-plane shown in the table. One also can convince oneself that, apart from the familiar combinations O3/O7 and O5/O9, this O5/O7 setup is the only other possibility of having two different kinds of O-planes that both extend along the four non-compact directions and still preserve some supersymmetry.

We see in table \ref{table:IIBOplanes} that one can always do a single T-duality along the 1- or 2-direction (which are the directions of the two real one-forms present in any six-dimensional SU(2)-structure manifold) to obtain a type IIA compactification with four O6-planes along the coordinate axis. However, after such a T-duality we generically have non-geometric Q-fluxes.

The organization of this paper is as follows. In section \ref{IIBcomp}, we give some more details on SU(2)-structure compactifications, their orientifolds and the effective 4D supergravity action and set up our notation. In section \ref{Tdual}, we also discuss the relation to IIA compactifications on SU(3)-structure manifolds with O6-planes via a formal T-duality transformation. In section \ref{susysolutions} we show the impossibility of supersymmetric Minkowski vacua in our setup and discuss some aspects of AdS vacua and the validity of the supergravity approximation. Section \ref{models} is devoted to several explicit examples that consist of coset spaces involving semisimple and U(1)-group factors as well as twisted tori. All these examples allow for a left-invariant SU(2)-structure and we can therefore use the left-invariant forms as expansion forms which yields a consistent effective 4D action. We compute the scalar potentials and show explicitly, in a simple example, how all moduli can be stabilized at tree-level. In section \ref{nogos}, we discuss the cosmological properties of our class of compactifications and derive several no-go theorems that forbid dS vacua and slow-roll inflation under certain assumptions. We then apply these no-go theorems to our explicit models and show that only one evades them. For that model we find a dS extremum with numerically vanishing $\epsilon$ and two tachyonic directions. A conclusion is given in section \ref{conclusion}. Finally, three appendices summarize some technical details.

\section{Type IIB compactifications on SU(2)-structure orientifolds}\label{IIBcomp}
In this section we discuss the $\mathcal{N}=1$ supergravity theory obtained from type IIB compactifications on an SU(2)-structure manifold with O5- and O7-planes. We also show that generically the resulting scalar potential is formally T-dual to a type IIA compactification on an SU(3)-structure space with O6-planes and non-geometric Q-fluxes.

\subsection{Manifolds with SU(2)-structure}
A six-dimensional manifold, $\mathcal{M}$, with (static) SU(2)-structure admits two globally defined and mutually orthogonal spinors $\eta$ and $\tilde{\eta}$, which we choose to be of unit norm. The existence of these two spinors means that the tangent space group can be restricted to  SU(2)$\subset$ SO(6). The spinors define a complex one-form, $V$, a real two-form, $\omega_2$, and a complex two-form, $\Omega_2$,
via suitable spinor bilinears on $\mathcal{M}$
\begin{eqnarray}
V_m &:=&  \frac{1}{2} \eta_{-}^{\dagger}\gamma_{m}\tilde{\eta}_+,\\
\omega_{2\, mn}&:=& i\eta_{+}^{\dagger}\gamma_{mn}\eta_{+}-i\tilde{\eta}_{+}^{\dagger}\gamma_{mn}\tilde{\eta}_{+},\\
\Omega_{mn}&:=& \tilde{\eta}_{+}^{\dagger}\gamma_{mn}\eta_{+},
\end{eqnarray}
where the subscript $\pm$ refers to the chirality of the 6D spinors, and the $\gamma_{m\ldots n}$ are the usual antisymmetrized gamma matrices. These forms are invariant under the tangent space group SU(2) and determine the metric on $\mathcal{M}$. Due to the Fierz identities and the assumed orthonormality of the spinors  $\eta$ and $\tilde{\eta}$,  they  satisfy a number of constraints,
\begin{eqnarray}
\omega_2\wedge \omega_2&=&\frac{1}{2}\Omega_2\wedge\Omega_2^{\ast}\neq0,\\
\omega_2\wedge\Omega_2&=&0,\quad\Omega_2\wedge\Omega_2=0,\\
\iota_{V}\Omega_2&=&0,\quad \iota_V\omega_2=0. \label{eq:iVoO}
\end{eqnarray}
A set of SU(2)-invariant forms with these properties provides an equivalent definition of an SU(2)-structure.

In the special case of SU(2)-holonomy (i.e., for $\mathcal{M}=T^2\times K3$), the two spinors are covariantly constant with respect to the Levi-Civita connection, which then implies that the forms $V,\omega_2,\Omega_2$ are all closed. On a general SU(2)-structure manifold, however, the spinors are not covariantly constant with respect to the Levi-Civita connection, and $V,\omega_2,\Omega_2$ are in general no
longer closed. Nevertheless, also in this generic case, one can define a different, torsionful, connection with respect to which the spinors are covariantly constant.
The nontrivial torsion of this connection is encoded in the non-vanishing exterior derivatives of the forms $V,\omega_2,\Omega_2$. Unlike the special case of a manifold with SU(2)-holonomy, a generic SU(2)-structure manifold is in general not Ricci-flat.


\subsection{Effective theories and field expansions}
In order to extract an effective 4D field theory from a given string compactification, one has to expand the higher-dimensional fields and fluxes in an appropriate set of expansion forms of the compact space. A 4D Lagrangian with finitely many fields requires the restriction to a finite set of such expansion forms. For this to be a meaningful theory, no interference with the neglected modes should spoil the dynamics of the modes one has kept, at least not in the regime the truncated theory is supposed to be valid. One way to ensure such a decoupling is a sufficiently large mass gap between the two sets of modes, as it may occur for example in  compactifications on Ricci-flat spaces such as tori or Calabi-Yau manifolds. There an expansion in terms of  harmonic forms provides the  classically massless moduli, well-separated from the massive Kaluza-Klein excitations.

On a generic SU(2)-structure manifold, however, the forms $V,\omega_2,\Omega_2$ are not closed and hence cannot be expanded in terms of harmonic forms. Metric deformations therefore tend to descend to massive 4D modes, and it is now less trivial to divide them into light and heavy fields \footnote{For a discussion of this issue see e.g. refs. \cite{KashaniPoor:2006si,KashaniPoor:2007tr}.}.

Another situation in which the restriction to finitely many fields is justified is when the neglected modes cannot be excited at all by the dynamics of the modes one has retained. This latter case is commonly referred to as ``consistent truncation" and means that any solution of the truncated 4D theory lifts to an exact solution of the full 10D theory.

The examples we discuss in detail in this paper are expected to be models in which a consistent truncation is possible. More precisely, we will consider in detail the case where $\mathcal{M}$ is a group manifold, or a quotient thereof by suitable discrete or continuous subgroups\footnote{The two classes we consider are: (i) products of compact semisimple and Abelian group factors, possibly modded out by suitable continuous subgroups (so as to yield  coset spaces), and (ii) nilmanifolds (or ``twisted tori''), i.e., nilpotent Lie-groups modded out by appropriate discrete subgroups.}. The  SU(2)-structure is furthermore required to respect this group structure, i.e., to be left-invariant under the group multiplication. The left-invariant forms on $\mathcal{M}$ are then taken as the natural expansion basis  that in some sense generalize the harmonic forms on a torus and lead to consistent truncations as argued in \cite{Scherk:1979zr,Hull:2005hk,Cassani:2009ck}.

The manifolds so-obtained are all parallelizeable, and would lead to an effective 4D theory with $\mathcal{N}=4$ supersymmetry.
In order to obtain a theory with $\mathcal{N}=1$ supersymmetry, we will  introduce two types of orientifold projections corresponding to O5- and O7-planes with orientations as in table \ref{table:IIBOplanes}. The left-invariant forms that survive these orientifold projections are labeled by $Y^{(npq)}$, where $n$ denotes the degree of the form and $p,q = \pm$ refer to the transformation property of the form under the O5- and O7-orientifold projection, respectively. In our examples these  forms exhibit a natural $(2+4)$-split \footnote{The complex vector field $V$ of a six-dimensional SU(2)-structure manifold defines a so-called almost product structure, i.e., a tensor field of rank $(1,1)$ that divides the tangent spaces into two-dimensional subspaces spanned by the real components of $V$ and well-defined four-dimensional complements. The two-forms $\omega_{2}$ and $\Omega_{2}$ of the SU(2)-structure have their legs stretched  only along  these complements (see e.g. \cite{Louis:2009dq,Bovy:2005qq}).}
   that is compatible with the orientation of our O-planes in table \ref{table:IIBOplanes} and in many ways parallels  properties of the $T^2\times K3$-compactification.  More precisely, in our expansion basis, there are two one-forms, and they can be only of the type $Y_{a}^{(1-+)}$ $(a=1,2)$ so that they have their leg only along the directions 1 and 2 in table \ref{table:IIBOplanes}. The two-forms, on the other hand, either arise as products of the two one-forms $Y_a^{(1-+)}$, or they have both legs along the remaining directions 3,4,5,6 in table \ref{table:IIBOplanes}. The expansion forms of rank higher than 2 all turn out to be obtainable from wedge products of the lower rank forms, so that the independent expansion forms are \footnote{In general it should also be possible to have compactifications with 2-forms along the four-dimensional part of the compact space that are even under both the O5- and the O7-orientifold projection. This would lead to D-terms as is discussed in detail in appendix \ref{app:DtermsIIB}. In the concrete examples we study in detail in this paper, however, a left-invariant $(++)$-two-form does not occur.}
\begin{center}
\begin{tabular}{ll}
0-form: & $Y^{(0++)}$, \\
1-forms: & $Y^{(1-+)}_a$,\\
2-forms: & $Y^{(2--)}_i$,\\
& $Y^{(2+-)}_A$,\\
& $Y^{(2-+)}_I,$
\end{tabular}
\end{center}
where the indices $i,j,\ldots$, $A,B,\ldots$, $I,J,\ldots$ label the  two-forms with the legs along the directions 3,4,5,6. Just as in the $T^2\times K3$-case, our setups have, up to multiplication by a function, only one four-form, $Y^{(4++)}$, with legs along the 3,4,5,6 directions. When wedged with the two one-forms $Y^{(1-+)}_a$, this form also yields the six-dimensional volume form of the full 6D space. We normalize it such that $\int Y^{(1-+)}_1 \wedge Y^{(1-+)}_2 \w  Y^{(4++)}=1$, which we often will also write as $\e_{ab} = \int Y^{(1-+)}_a \wedge Y^{(1-+)}_b \wedge Y^{(4++)}$ with $\e_{12} = -\e_{21}=1$.\\
Products of two-forms with different orientifold parities can never combine to an even/even four-form,  and, due to the uniqueness of $Y^{(4++)}$, thus have to vanish. $Y^{(4++)}$ can therefore only be obtained from products of two-forms of the same parity, and we only have the following non-vanishing symmetric intersection forms
\ba
\tilde{X}_{ij} &=& \int Y^{(1-+)}_1 \wedge Y^{(1-+)}_2 \wedge Y^{(2--)}_i \wedge Y^{(2--)}_j \nn, \\
\hat{X}_{AB} &=& \int Y^{(1-+)}_1 \wedge Y^{(1-+)}_2 \wedge Y^{(2+-)}_A \wedge Y^{(2+-)}_B, \\
\bar{X}_{IJ} &=& \int Y^{(1-+)}_1 \wedge Y^{(1-+)}_2 \wedge Y^{(2-+)}_I \wedge Y^{(2-+)}_J. \nn
\ea
Before we expand the fields and fluxes in the above-described expansion forms, we summarize their transformation properties under O5- and O7-orientifold projections \cite{Koerber:2007hd}.
\begin{center}
\begin{tabular}{|c|c|c|}
  \hline
  Fields & O5 & O7\\ \hline
  $C_0, F_1$ & - & +\\ \hline
  $C_2, F_3$ & + & -\\ \hline
  $C_4, F_5$ & - & +\\ \hline
  $C_6$      & + & -\\ \hline
  $B, H_3$ & - & -\\ \hline
  $\omega_2$ & - & -\\ \hline
  Re$(\Omega_2)$ & - & +\\ \hline
  Im$(\Omega_2)$ & + & -\\ \hline
  V & - & +\\ \hline
\end{tabular}
\end{center}
Since the 0-form and the volume form on the six-dimensional space are both even under the O5- and the O7-orientifold projections we find that the RR-axions $C_0$ and $C_6$ are projected out. (This is required since, as we will argue below, after one formal T-duality we have a manifold without 1- and 5-forms and therefore without $C_1$ and $C_5$.) The scalar components for the remaining fields can be combined into complex fields and expanded as follows \cite{Koerber:2007hd}, \cite{Caviezel:2008ik}
\ba\label{eq:fields}
\om^c &=& \om_2 - i B = (k^i - i b^i) Y^{(2--)}_i = t^i Y^{(2--)}_i, \nn\\
\Om^c_2 &=& e^{-\phi} \text{Im}(\Om_2) + i C_2 = (u^A + i c_{(2)}^A) Y^{(2+-)}_A = z^A Y^{(2+-)}_A, \nn\\
\Om^c_4 &=& -i e^{-\phi} 2 V \wedge V^* \wedge \text{Re}(\Omega_2) + i C_4 = (v^I + i c_{(4)}^I) Y^{(1-+)}_1 \wedge Y^{(1-+)}_2 \wedge Y^{(2-+)}_I  \\
&=& w^I Y^{(1-+)}_1 \wedge Y^{(1-+)}_2 \wedge Y^{(2-+)}_I, \nn\\
2 V &=& L \lp i Y^{(1-+)}_1 - (x+i y) Y^{(1-+)}_2 \rp = L \lp i Y^{(1-+)}_1 - \tau Y^{(1-+)}_2 \rp = L\, T^a Y^{(1-+)}_a. \nn
\ea
In the last line we have only one complex modulus $\tau$, since the overall scale, $L$, of $2V$ drops out of the scalar potential and can be eliminated from the K\"ahler and superpotential through a K\"ahler transformation. We do not get any vector fields from the metric or the $B$ field, since the 1-forms $Y^{(1-+)}_a$ are odd/even under the O5- and O7-orientifold projections, and the metric is even and the $B$ field odd under both projections. \\
Next we expand the background fluxes in our basis
\eq{\label{eq:fluxes}\spl{
F_1 &= m^a Y^{(1-+)}_a,  \\
F_3 &= e^{ai} Y^{(1-+)}_a \wedge Y^{(2--)}_i,  \\
F_5 &= f^{a} Y^{(1-+)}_a \wedge Y^{(4++)},  \\
H_3 &= h^{aA} Y^{(1-+)}_a \wedge Y^{(2+-)}_A.
}}
As mentioned above, in a generic SU(2)-structure compactification, the forms $V,\omega_2,\Omega_2$ are in general not closed, so that we have   to  allow for the possibility of having non-closed expansion forms, i.e. a deviation from the SU(2)-holonomy case, which we parameterize as follows
\ba\label{eq:IIBd}
\dd Y^{(1-+)}_a &=& r_a^I Y^{(2-+)}_I, \nn \\
\dd Y^{(2--)}_i &=& \tilde{r}^{aA}_i Y^{(1-+)}_a \wedge Y^{(2+-)}_A, \nn \\
\dd Y^{(2+-)}_A &=& \hat{r}^{ai}_A Y^{(1-+)}_a \wedge Y^{(2--)}_i, \\
\dd Y^{(2-+)}_I &=& 0. \nn
\ea
Here, the coefficients $r_a^I$, $\tilde{r}_i^{aA}$ and $\hat{r}_A^{ai}$ are constant parameters. \\
Note that the matrices $\tilde{r}^a$ and $\hat{r}^a$ are not independent, as we have \footnote{Here and at various other places in the following, we use that $\dd Y_a^{(1-+)}\propto Y_{I}^{(2-+)}$, whose wedge product with $Y_i^{(2--)}$ or $Y_A^{(2+-)}$ vanishes, so that there is no contribution from $\dd$ acting on $ Y_a^{(1-+)}$.}
\ba
- \e_{ab} \hat{X}_{AB} \tilde{r}_i^{bB} &=& \int \dd Y^{(2--)}_i \w Y_a^{(1-+)}\w Y_A^{(2+-)} \\
&=& \int Y^{(2--)}_i \w Y_a^{(1-+)} \w \dd Y_A^{(2+-)} = \e_{ab} \tilde{X}_{ij} \hat{r}_A^{bj} \nn,
\ea
so that
\begin{equation}\label{hattilde}
\hat{r}_A^{ai} = -\hat{X}_{AB} \left( \tilde{X}^{-1} \right)^{ij} \tilde{r}_j^{aB}.
\end{equation}
We will nevertheless use both $\hat{r}_A^{ai}$ and $\tilde{r}_j^{aB}$ in explicit formulas to simplify expressions. The reader should keep in mind, though, that they are not independent and in particular, if one of them vanishes, so does the other.\\
Demanding that $\dd$ squares to zero on the forms \footnote{This is a necessary but not sufficient condition one has to impose on the metric fluxes. The sufficient condition is given below in \eqref{eq:Bianchi}.} one furthermore finds
\be\label{eq:dsquaredIIB}
\e_{ab} \tilde{r}_i^{aA} \hat{r}_A^{bj} = \e_{ab} \hat{r}_A^{ai} \tilde{r}_i^{bB} = 0.
\ee
The RR-fluxes given in \eqref{eq:fluxes} are constrained by the Bianchi identities that have the following form
\ba\label{eq:tadpole}
\dd F_1 &=& m^a r_a^I Y_I^{(2-+)} = 4 [\delta_{O7}], \nn \\
\dd F_3 + H_3 \w F_1 &=& \left( -e^{ai} \tilde{r}_i^{bA} + h^{aA} m^b \right) Y_a^{(1-+)} \w Y_b^{(1-+)} \w Y_A^{(2+-)} = [\delta_{O5}], \\
\dd F_5 + H_3 \w F_3 &=& 0 = \frac{1}{4} [\delta_{O3}], \nn
\ea
where $[\delta_{Op}]$ denotes the $(9-p)$-form contribution of the smeared Op-planes. We see that in the absence of D-branes the setup only allows for O5- and O7-planes but no O3-planes.\\
The absence of NS5-branes finally requires the closure of the $H_3$ flux
\be\label{eq:dH}
\dd H_3 = -\e_{ab} h^{aA} \hat{r}_A^{bi} \, Y_1^{(1-+)} \w Y_2^{(1-+)} \w Y_i^{(2--)} =0,
\ee
which gives the extra constraint
\begin{equation}
\e_{ab} h^{aA} \hat{r}_A^{bi}=0.
\end{equation}

\subsection{The four-dimensional scalar potential}
With the conventions given above we can now discuss the resulting four-dimensional theory. We started with 32 real supercharges in type IIB, and the compactification on the SU(2)-structure manifold would preserve only half of this original supersymmetry in the resulting effective action. The two orientifold projections break each another half so that we are left with an effective 4D action with $\frac{1}{8}$ of the original supersymmetry, i.e. four real supercharges corresponding to $\mathcal{N}=1$. Note that any two of the orientifold planes in our setup of table \ref{table:IIBOplanes} have four Neumann-Dirichlet directions and thus can preserve a common set of supercharges \footnote{Strictly speaking our setup is an asymmetric orbifold $T^6/(\mathbb{Z}_2 \times (-1)^{F_L} \mathbb{Z}_2)$ with a single orientifold projection. The generators of the orbifold group together with the single orientifold projection then give rise to both O5- and O7-planes. Since our compact space is the special type of asymmetric orbifold given above, it should still be possible to use supergravity. We thank Ralph Blumenhagen for bringing this issue to our attention.}. We therefore can write the four-dimensional action using the language of $\mathcal{N} =1$ supergravity. Our main interest is in the scalar potential for the closed string moduli that is determined by the K\"ahler potential $K$ and the superpotential $W$ as \footnote{See appendix \ref{app:DtermsIIB} for the four-dimensional action and potential D-term contributions to the scalar potential.}
\be\label{eq:scalarpotential}
V = e^K\lp K^{M\bar N}D_MW\,\overline{D_{ N} W}-3|W|^2\rp.
\ee
To determine the K\"ahler potential $K$ and the superpotential $W$ we can plug our expansions from the previous subsection into the generic expressions for the K\"ahler and superpotential for  SU(3)$\times$SU(3) structure compactifications \cite{Grimm:2004uq,Grana:2005ny,Grana:2006hr,Benmachiche:2006df,Koerber:2007hd,Koerber:2007xk,Cassani:2007pq,Cassani:2008rb}. For the K\"ahler potential we find
\ba\label{eq:Kaehlerpotential}
K &=& K_k + K_{cs}, \nn \\
K_k &=& - \ln \ls \frac{i}{|L|^2} \int \langle 2 V \w  e^{i\om_2}, 2 V^* \w e^{-i \om_2} \rangle \rs = - \ln \ls \frac{-2i}{|L|^2} \int 2 V \w 2 V^* \w \om_2 \w \om_2 \rs \nn\\
&=& - \ln \ls -(\tau + \bar{\tau})\hlf\tilde{X}_{ij} (t^i+\bar{t}^i)(t^j+\bar{t}^j) \rs, \\
K_{cs} &=& - 2 \ln \ls \frac{i}{8} \int \langle e^{-\phi} e^{2 V\w V^*} \Om_2,\overline{e^{-\phi} e^{2 V\w V^*} \Om_2} \rangle \rs \nn\\
&=& - 2\ln \ls -\frac{i}{2} \int \left( e^{-2\phi} V \w V^* \w \Om_2 \w \Om_2^*\right) \rs \nn\\
&=& - 2\ln \ls -i \int \left( e^{-2\phi} V \w V^* \w \om_2 \w \om_2 \right) \rs \nn \\
&=& - 2 \ln \ls e^{-2D}\rs = 4 D,\nn
\ea
where $\langle \, ,\rangle$ is the Mukai pairing whose action on polyforms $A,B$ is given by $\langle A,B \rangle = A \w \varpi(B)|_{6-form}$. The operator $\varpi$ acts on forms by inverting the order of its coordinate indices, and $|_{6-form}$ means that we keep only the six-form part. We also used the explicit expansion of the fields given in \eqref{eq:fields}. In the second to last step we used the SU(2)-structure condition $\om_2 \w \om_2 =\frac{1}{2} \Om_2 \w \Om_2^*$ and in the last step we introduced the four-dimensional dilaton $D$ that satisfies $e^{-2D} = e^{-2 \phi} vol_6$. We have written $K_{cs}$ in terms of the four-dimensional dilaton to facilitate the discussion of the formal T-duality below. For the explicit form of $K_{cs}$ in terms of the complex structure moduli see \eqref{eq:Kaehlerpotential2}. The volume of the compact space is given by $vol_6 = -i \int V \w V^* \w \om_2 \w \om_2 = -|L|^2 x \frac{1}{2} \tilde{X}_{ij} k^i k^j$ \footnote{The value of $|L|^2$ is fixed through equation \eqref{eq:SU2condition} to be $|L|^2 = \frac{1}{x} \sqrt{\frac{\bar{X}_{IJ} v^I v^J}{\hat{X}_{AB} u^A u^B}}$ so that we have $vol_6 = -\sqrt{\frac{\bar{X}_{IJ} v^I v^J}{\hat{X}_{AB} u^A u^B}} \frac{1}{2} \tilde{X}_{ij} k^i k^j$.}, where in our conventions the K\"ahler moduli are all positive $x, k^i>0$ and the intersection number $\tilde{X}_{ij}$ will have negative entries.\\
In terms of the polyforms $F=\sum F_p$ and $C = \sum C_p$ the superpotential is given by
\be
W = -\frac{i}{2 L}\int \langle 2V \w e^{i (\om_2-i B)}, F -i \lp \dd + H_3\w \rp \lp e^B e^{-\phi} \text{Im}(e^{2V \w V^*} \w \Om_2)+i C \rp \rangle.
\ee
The parity assignments  of  $B$ and $\Omega_2$ imply that their wedge product cannot combine to the unique four-form $Y^{(4++)}$ and hence must vanish, so that we have $e^B \w \Om_2 = \Om_2$. In terms of the complex fields \eqref{eq:fields} the superpotential therefore becomes
\ba\label{eq:superpotential}
&&W =-\frac{i}{2 L}  \int \langle 2V \w e^{i \om^c},F -i (\dd + H_3\w) (\Om^c_2 + \Om^c_4) \rangle \\
&=&  -\frac{i}{2 L}\int \langle 2V + 2i V \w \om^c - V \w \om^c \w \om^c,F_1 + \lp F_3 - i \dd \Om^c_2 \rp + \lp F_5 - i H_3 \w \Om^c_2  -i \dd \Om^c_4 \rp \rangle  \nn\\
&=&  -\frac{i}{2} T^a \lp \e_{ab} \ls f^b - i \hat{X}_{AB} h^{bA} z^B - \tilde{X}_{ij} t^i \lp i e^{bj} + z^A \hat{r}_A^{bj} +\hlf m^b t^j\rp \rs -i \bar{X}_{IJ} w^I r_a^J \rp. \nn
\ea

\subsection{T-duality to type IIA on SU(3)-structure manifolds}\label{Tdual}
We now proceed by showing that the above scalar potential is formally T-dual to the scalar potential obtained by compactifying type IIA on an SU(3)-structure space with non-geometric fluxes (so called Q-fluxes \cite{Wecht:2007wu}). The resulting K\"ahler and superpotential for such compactifications are \cite{Shelton:2005cf,Robbins:2007yv}
\ba\label{eq:IIApotentials}
K_{(IIA)} &=& - \ln \ls \frac{1}{6} \kappa_{abc} (t^a_{(IIA)}+\bar{t}^a_{(IIA)})(t^b_{(IIA)}+\bar{t}^b_{(IIA)})(t^c_{(IIA)}+\bar{t}^c_{(IIA)}) \rs \nn \\
&& + 4 D_{(IIA)},\\
W_{(IIA)} &=& -\frac{i}{2} \ls -f^{(6)}+ i t^a_{(IIA)} f^{(4)}_a + \hlf \k_{abc} t^a_{(IIA)} t^b_{(IIA)} f^{(2)c} - \frac{i}{6} f^{(0)} \k_{abc}t^a_{(IIA)} t^b_{(IIA)} t^c_{(IIA)} \right. \nn\\
&& \left. + i h^{(3)}_K Z^K_{(IIA)} + r_{aK} t^a_{(IIA)} Z^K_{(IIA)}- \frac{i}{2} \k_{abc} q^a_K t^b_{(IIA)} t^c_{(IIA)} Z^K_{(IIA)} \rs,
\ea
where $f^{(p)}$ denote the RR-fluxes, $h^{(3)}_K$ the $H_3$-flux, $r_{aK}$ the metric fluxes and $q^a_K$ the non-geometric fluxes. $t^a_{(IIA)}$ are the K\"ahler moduli and $Z^K_{(IIA)}$ are the complex structure moduli. \\
We can formally T-dualize the type IIB models along either $Y^{(1-+)}_1$ or $Y^{(1-+)}_2$ to go to type IIA. The T-duality along $Y^{(1-+)}_1$ leaves all the moduli invariant, and we can reinterpret the type IIB K\"ahler and superpotential as type IIA superpotential arising from a compactification on an SU(3)-structure manifold. After a K\"ahler transformation $W_{(IIA)} \rightarrow -i W_{(IIA)}$ that leaves the K\"ahler potential invariant, one can compare the K\"ahler and superpotential and finds that they are identical if one identifies
\ba\label{eq:T1}
t^a_{(IIA)} &=& (\tau,t^i), \quad Z^K_{(IIA)} = \left(\begin{array}{c} z^A \\ w^I \\ \end{array}\right), \quad D_{(IIA)} =D, \nn\\
f^{(6)} &=& f^2,\quad f^{(4)}_a = (f^1,\tilde{X}_{ij}e^{2j}),\quad f^{(2)a} =(-m^2,-e^{1i}),\quad f^{(0)} =-m^1,\\
h^{(3)}_K &=& \left(\begin{array}{c} \hat{X}_{AB} h^{2B} \\ \bar{X}_{IJ} r_1^J \\ \end{array}\right), \quad r_{aK} =\left(\begin{array}{cc} \hat{X}_{AB} h^{1B} & \tilde{X}_{ij} \hat{r}^{2j}_A \\ -\bar{X}_{IJ} r_2^J & 0 \\ \end{array} \right), \quad q^a_K =\left(\begin{array}{cc} 0 & -\hat{r}^{1j}_A \\ 0 & 0 \\ \end{array} \right),\nn
\ea
and the only non-vanishing components of the symmetric triple intersection number are $\kappa_{1ij}=-\tilde{X}_{ij}$. This all agrees nicely with the T-duality rules of \cite{Buscher:1987sk,Hassan:1999mm} and the rule \cite{Shelton:2005cf,Wecht:2007wu},  which state that generalized NSNS-fluxes with no leg along the T-duality direction are invariant while the ones with a leg along the T-duality direction transform into other types of generalized NSNS-fluxes. In particular we see that models for which $\hat{r}^{1i}_A \neq 0$ are formally T-dual to SU(3)-structure compactifications with non-geometric fluxes.\\
For a T-duality along $Y^{(1-+)}_2$ we find that the K\"ahler modulus $\tau$ in front of $Y^{(1-+)}_2$ gets inverted $\tau \rightarrow \tau'= \frac{1}{\tau}$. We transform the superpotential $W \rightarrow \tau' W$ and the K\"ahler potential $K \rightarrow K - \ln (\tau') -\ln (\bar{\tau}')$. This results in a type IIB K\"ahler potential as given in \eqref{eq:Kaehlerpotential} with $\tau$ replaced by $\tau'$ and the superpotential is as given in \eqref{eq:superpotential} where the only change is that $T^a =(i,-\tau) \rightarrow (i\tau',-1)$. After this transformation we can again identify our superpotential and K\"ahler potential as arising from a compactification of type IIA on an SU(3)-structure space with non-geometric fluxes. In particular we find that they agree, if we make the following identifications
\ba
t^a_{(IIA)} &=& (\tau',t^i), \quad Z^K_{(IIA)} = \left(\begin{array}{c} z^A \\ w^I \\ \end{array}\right), \quad D_{(IIA)} =D, \nn\\
f^{(6)} &=&-f^1, \quad f^{(4)}_a = (f^2,-\tilde{X}_{ij} e^{1j}),\quad f^{(2)a} =(m^1,-e^{2i}),\quad f^{(0)} =-m^2,\\
h^{(3)}_K &=& \left(\begin{array}{c} -\hat{X}_{AB} h^{1B} \\ \bar{X}_{IJ} r_2^J \\ \end{array}\right), \quad r_{aK} =\left(\begin{array}{cc} \hat{X}_{AB} h^{2B} & -\tilde{X}_{ij} \hat{r}^{1j}_A \\ \bar{X}_{IJ} r_1^J & 0 \\ \end{array} \right), \quad q^a_K =\left(\begin{array}{cc} 0 & -\hat{r}^{2i}_A \\ 0 & 0 \\ \end{array} \right),\nn
\ea
where again the only non-vanishing components of the symmetric triple intersection number are $\kappa_{1ij}=-\tilde{X}_{ij}$.\\
This shows that a formal T-duality along either of the two 1-cycles corresponding to the two 1-forms leads to a compactification on type IIA on a space with SU(3)-structure that is generically only locally geometric. In particular, SU(2)-structure compactifications for which $\hat{r}^{1i}_A \neq 0$ and $\hat{r}^{2i}_A \neq 0$ are not T-dual to any geometric compactification of type IIA on SU(3)-structure manifolds.

\section{Analysis of supersymmetric vacua of type IIB on SU(2)-structure compactifications}\label{susysolutions}

In this section we discuss the existence of supersymmetric vacua in compactifications of type IIB theory on SU(2)-structure manifolds in the presence of O5- and O7-planes. We study the F-term equations arising from the K\"ahler potential \eqref{eq:Kaehlerpotential} and superpotential \eqref{eq:superpotential}. While there are no obstructions in  finding fully stabilized AdS vacua, it is not possible to stabilize all moduli in supersymmetric Minkowski vacua. We also discuss consistency conditions like the tadpole condition and the possibility of obtaining large volume and small string coupling in these compactifications.

\subsection{Supersymmetric Minkowski vacua}
In \cite{Micu:2007rd,Ihl:2007ah} the authors show that for geometric compactifications of type IIA on SU(3)-structure spaces it is not possible to stabilize all moduli in a supersymmetric Minkowski vacuum. However, in \cite{Micu:2007rd} the authors argue that with non-geometric fluxes it is possible to find supersymmetric Minkowski vacua. Since our type IIB compactifications are formally T-dual to certain type IIA compactifications on SU(3)-structure spaces with non-geometric fluxes the question of whether they allow for fully stabilized supersymmetric Minkowski vacua is of obvious interest.\\
In order to find supersymmetric Minkowski vacua we have to find solutions for which the superpotential \eqref{eq:superpotential}
\be\label{eq:superpotential2}
W = -\frac{i}{2} T^a \lp \e_{ab} \ls f^b - i \hat{X}_{AB} h^{bA} z^B - \tilde{X}_{ij} t^i \lp i e^{bj} + z^A \hat{r}_A^{bj} +\hlf m^b t^j\rp \rs -i \bar{X}_{IJ} w^I r_a^J \rp,
\ee
and its derivatives with respect to all the moduli vanish. This means that we have to solve an over-determined system of equations since we have one complex equation for every complex modulus plus the extra complex equation $W=0$. So for generic fluxes one expects no solution \footnote{There is always the solution in which all moduli are zero. We neglect this trivial solution, which corresponds to a compact space with zero volume, since our supergravity analysis is certainly not applicable in this case.}. However, one can hope that for special values of the fluxes it is possible to find solutions that stabilize all moduli. Recalling that $T^a = (i,-\tau)$ the equations for supersymmetric Minkowski vacua are
\ba
0 &=& W,\\
0 &=& \p_\tau W = -\frac{i}{2} \lp f^1 - i \hat{X}_{AB} h^{1A} z^B - \tilde{X}_{ij} t^i \lp i e^{1j} + z^A \hat{r}_A^{1j} +\hlf m^1 t^j\rp  + i \bar{X}_{IJ} w^I r_2^J \rp, \quad {}\\
0 &=& \p_{t^i} W = \frac{i}{2} T^a \e_{ab} \tilde{X}_{ij} \lp i e^{bj} + z^A \hat{r}_A^{bj} + m^b  t^j \rp,\\
0 &=& \p_{z^A} W = \frac{i}{2} T^a \e_{ab} \lp i \hat{X}_{AB} h^{bB} + \tilde{X}_{ij} t^i \hat{r}_A^{bj} \rp,\\
0 &=& \p_{w^I} W = -\frac{1}{2} \bar{X}_{IJ} T^a r_a^J = \hlf \tau \bar{X}_{IJ} r_2^J - \frac{i}{2} \bar{X}_{IJ} r_1^J.
\ea
Taking the real part of the last equation and using that the K\"ahler modulus $x >0$ and that the intersection number $\bar{X}_{IJ}$ is invertible we find
\be
0 = 2 \text{Re}(\p_{w^I} W) = x \bar{X}_{IJ} r_2^J \quad \Rightarrow \quad r_2^J =0, \, \forall J.
\ee
This then implies that $0 = 2 i \p_{w^I} W = \bar{X}_{IJ} r_1^J$ and we can conclude that only manifolds that have $r_a^J=0, \forall a,J$ can potentially have Minkowski vacua. Furthermore, we see from the superpotential \eqref{eq:superpotential2} that in that case the superpotential does not depend on the $w^I$ which means that any supersymmetric Minkowski vacuum will always have the $w^I$ as flat directions. So we can conclude that there are no fully stabilized supersymmetric Minkowski vacua possible in these kind of compactifications.\\
Nevertheless, one could proceed to analyze the F-term equations further and check how many moduli one can actually stabilize. However, from the tadpole condition \eqref{eq:tadpole} we see that $r_a^J=0 \, \Rightarrow \, \dd F_1 = 0 = 4 [\delta_{O7}]$. This means that it is impossible to satisfy this condition in our setup. A way around this would be to cancel the O7-plane charge using D7-branes. However, this would lead to open string moduli associated with the D7-branes, which we do not consider in this paper. Another possibility is not to do the orientifold projection that leads to the O7-planes. This should give a four-dimensional $\mathcal{N}=2$ theory. Both of these possibilities are beyond the scope of this paper and we do not explore them any further.

\subsection{Supersymmetric AdS vacua}
In order to calculate the F-term equations we need the derivatives of the K\"ahler potential \eqref{eq:Kaehlerpotential} and therefore would like to have an explicit expression of the K\"ahler potential in the complex structure sector. We can get this by using the SU(2)-structure condition
\ba\label{eq:SU2condition}
0 &=& \Om_2 \w \Om_2 = \text{Re} (\Om_2) \wedge \text{Re} (\Om_2) - \text{Im} (\Om_2) \wedge \text{Im} (\Om_2) \nn \\
 &\Rightarrow& \quad \text{Re} (\Om_2) \wedge \text{Re} (\Om_2) = \text{Im} (\Om_2) \wedge \text{Im} (\Om_2),
\ea
where the mixed term $\text{Re} (\Om_2) \wedge \text{Im} (\Om_2)$ vanishes since there is no odd/odd four-form. We therefore have
\be
\Om_2 \w \Om_2^* = 2 \text{Re} (\Om_2) \wedge \text{Re} (\Om_2) = 2 \text{Im} (\Om_2) \wedge \text{Im} (\Om_2).
\ee
Using the explicit expansion of the fields as given in \eqref{eq:fields} we can rewrite the K\"ahler potential in the complex structure sector as an explicit function of the complex moduli in the complex structure sector (we assume constant dilaton $\phi$)
\ba\label{eq:Kaehlerpotential2}
K_{cs} &=& - 2\ln \ls -\frac{i}{8} \int \left( e^{-2\phi} 2 V \w 2V^* \w \Om_2 \w \Om_2^*\right) \rs \nn\\
&=& - 2\ln \ls -\frac{|L|^2 x}{2}e^{-2\phi} \int \left( Y_1^{(1-+)} \w Y_2^{(1-+)} \w \text{Re} (\Om_2) \wedge \text{Re} (\Om_2) \right) \rs \nn \\
&=& - \ln \ls -\frac{|L|^2 x}{2}e^{-2\phi} \int \left( Y_1^{(1-+)} \w Y_2^{(1-+)} \w \text{Re} (\Om_2) \wedge \text{Re} (\Om_2) \right) \rs \nn \\
&&- \ln \ls -\frac{|L|^2 x}{2}e^{-2\phi} \int \left( Y_1^{(1-+)} \w Y_2^{(1-+)} \w \text{Im} (\Om_2) \wedge \text{Im} (\Om_2) \right) \rs\\
&=& - \ln \ls -\frac{|L|^2 x}{2} \lp \frac{1}{|L|^4 x^2} \bar{X}_{IJ} v^I v^J \rp \rs - \ln \ls -\frac{|L|^2 x}{2} \hat{X}_{AB} u^A u^B \rs \nn \\
&=& - \ln \ls \hlf \hat{X}_{AB} u^A u^B \hlf \bar{X}_{IJ} v^I v^J \rs \nn \\
&=& - \ln \ls \frac{1}{64}\hat{X}_{AB} (z^A+\bar{z}^A) (z^B+\bar{z}^B) \bar{X}_{IJ} (w^I+\bar{w}^I)(w^J+\bar{w}^J) \rs. \nn
\ea
Using this explicit form for the K\"ahler potential we can spell out the F-term equations $DW= \p W + W \p K=0$ for supersymmetric AdS vacua \footnote{For some recent discussion of non-supersymmetric AdS-vacua, see e.g., \cite{Lust:2008zd,Koerber:2010rn}.}
\ba\label{eq:Fterms}
0 &=& D_\tau W \nn \\
&=& -\frac{i}{2} \lp f^1 - i \hat{X}_{AB} h^{1A} z^B - \tilde{X}_{ij} t^i \lp i e^{1j} + z^A \hat{r}_A^{1j} +\hlf m^1 t^j\rp + i \bar{X}_{IJ} w^I r_2^J \rp - \frac{1}{2x}W, \nn \\
0 &=& D_{t^i} W = \frac{i}{2} T^a \e_{ab} \tilde{X}_{ij} \lp i e^{bj} + z^A \hat{r}_A^{bj} + m^b t^j \rp - \frac{\tilde{X}_{ij}k^j}{\tilde{X}_{kl}k^kk^l}W, \\
0 &=& D_{z^A} W = \frac{i}{2} T^a \e_{ab} \ls i \hat{X}_{AB} h^{bB} + \tilde{X}_{ij} t^i \hat{r}_A^{bj} \rs - \frac{\hat{X}_{AB}u^B}{\hat{X}_{CD}u^Cu^D}W, \nn\\
0 &=& D_{w^I} W = -\hlf \bar{X}_{IJ} T^a r_a^J - \frac{\bar{X}_{IJ}v^J}{\bar{X}_{KL}v^Kv^L}W = \hlf \tau \bar{X}_{IJ} r_2^J -\frac{i}{2} \bar{X}_{IJ} r_1^J - \frac{\bar{X}_{IJ}v^J}{\bar{X}_{KL}v^Kv^L}W.\nn
\ea

\subsection{The tadpole condition and the validity of the supergravity approximation}
The fluxes in the superpotential are not all independent but have to satisfy several constraints. As mentioned above in equations \eqref{eq:tadpole} there are tadpole conditions corresponding to the O5- and O7-planes as well as NS5-branes
\ba\label{eq:tadpole2}
\dd F_1 &=& m^a r_a^I Y_I^{(2-+)} = 4 [\delta_{O7}], \nn \\
\dd F_3 + H_3 \w F_1 &=& \left( -e^{ai} \tilde{r}_i^{bA} + h^{aA} m^b \right) Y_a^{(1-+)} \w Y_b^{(1-+)} \w Y_A^{(2+-)} = [\delta_{O5}], \\
\dd H_3 &=& -\e_{ab} h^{aA} \hat{r}_A^{bi} \, Y_1^{(1-+)} \w Y_2^{(1-+)} \w Y_i^{(2--)} =0.\nn
\ea
Above in \eqref{eq:dsquaredIIB} we also derived the constraints arising from demanding that $\dd^2=0$ when acting on our basis forms. However, the metric fluxes need to satisfy a stronger constraint. On a parallelizable manifold we have a set of globally defined one-forms $e^n,\, n=1,\ldots 6$. The metric fluxes are then defined by $\dd e^n =-\frac12 f^n_{mp} e^m \wedge \e^p$ and from demanding that $\dd^2=0$ on the one-forms we find
\be\label{eq:Bianchi}
f^n_{[mp} f^r_{s] n} = 0,
\ee
which in particular implies that $\dd^2 =0 $ when acting on our expansion forms. Note however, that \eqref{eq:Bianchi} is generically not implied by \eqref{eq:dsquaredIIB}.\\
In equation \eqref{eq:fluxes} we have expanded the $H_3$- and RR-fluxes in forms that are not all in cohomology. Since one can always shift the B- and RR-axions to set any exact part of the fluxes to zero, we have generically redundant parameters in our general superpotential \eqref{eq:superpotential2}. Explicitly, if we make the constant shift in the axions
\ba
c_{(2)}^A &\rightarrow& c_{(2)}^A + \delta c_{(2)}^A,\nn \\
c_{(4)}^I &\rightarrow& c_{(4)}^I + \delta c_{(4)}^I,\\
b^i &\rightarrow& b^i + \delta b^i, \nn
\ea
then this leads to the following changes in the fluxes
\ba
F_3 &=& e^{ai} Y^{(1-+)}_a \wedge Y^{(2--)}_i + \dd \lp c_{(2)}^A Y^{(2+-)}_A \rp \nn \\
&\rightarrow& \lp e^{ai}+\hat{r}_A^{ai} \delta c_{(2)}^A \rp Y^{(1-+)}_a \wedge Y^{(2--)}_i + \dd \lp c_{(2)}^A Y^{(2+-)}_A \rp,\nn \\
F_5 &=& f^{a} Y^{(1-+)}_a \wedge Y^{(4++)} + \dd\lp c_{(4)}^I Y^{(1-+)}_1 \wedge Y^{(1-+)}_2 \wedge Y^{(2-+)}_I \rp \\
&\rightarrow& \lp f^{a} + \e^{ab} \bar{X}_{IJ} r_b^J \delta c_{(4)}^I \rp Y^{(1-+)}_a \wedge Y^{(4++)} + \dd\lp c_{(4)}^I Y^{(1-+)}_1 \wedge Y^{(1-+)}_2 \wedge Y^{(2-+)}_I \rp, \nn \\
H_3 &=& h^{aA} Y^{(1-+)}_a \wedge Y^{(2+-)}_A + \dd\lp b^i Y^{(2--)}_i \rp\nn\\
&\rightarrow& \lp h^{aA} + \tilde{r}_i^{aA} \delta b^i \rp Y^{(1-+)}_a \wedge Y^{(2+-)}_A +\dd\lp b^i Y^{(2--)}_i \rp.\nn
\ea
We see that we can choose the $\delta b^i$ to set any exact part of the $H_3$-flux to zero and $\delta c_{(2)}^A$ and $\delta c_{(4)}^I$ for the exact parts of $F_3$ and $F_5$, respectively.\\
As long as we are discussing a generic compactification it is not possible to say which parts of the $H_3$- and the RR-fluxes are exact and which metric fluxes are non-vanishing so that the generic superpotential \eqref{eq:superpotential2} cannot be simplified. However, for any concrete model with metric fluxes there will be flux parameters that can be set to zero by shifting an axion. In the concrete models of the next section we will always set the exact part of the $H_3$- and RR-fluxes to zero.\\
We can only neglect corrections to our supergravity analysis when the volume of the compactification space is large and the string coupling is small. While for generic flux compactifications of type IIA on Calabi-Yau manifolds there exists a limit of large $F_4$-flux that leads simultaneously to large volume and small string coupling \cite{DeWolfe:2005uu}, we are not aware of any such generic statement in compactifications of type IIA on SU(3)-structure manifolds. This is certainly an interesting question whose answer should translate to our type IIB compactifications on SU(2)-structure manifolds. Rather than pursuing this question we will content ourselves with pointing out that the tadpole constraints \eqref{eq:tadpole2} do not involve the $F_5$-flux since the last equation in \eqref{eq:tadpole} is automatically satisfied. Furthermore, as we will see in an explicit example below, not all the flux parameters are necessarily constrained by the tadpole conditions. So one generically expects to have unconstrained fluxes and can hope to use these to obtain a large volume and a small string coupling so that the supergravity analysis is valid. We will demonstrate that this is possible in an explicit example in \ref{SU(3)U(1)}.

\section{Explicit Models}\label{models}

In this section we discuss several explicit examples of the SU(2)-structure compactifications introduced in the previous sections. We start out by analyzing compactifications of type IIB supergravity on cosets models \cite{Koerber:2008rx,Caviezel:2008ik} with SU(2)-structure and O5- and O7-planes. Then we discuss compactifications on spaces obtained from a base-fiber splitting \cite{Dabholkar:2005ve,Ihl:2007ah}. In all concrete models we restrict ourselves to only bulk moduli and fluxes for simplicity\footnote{For a discussion of blow-up moduli in IIA flux compactifications, see e.g. \cite{DeWolfe:2005uu}.}. For concreteness, we will choose the following expansion forms,
\begin{table}[h]
\begin{center}
\begin{tabular}{|l|ll|}
\hline
1-forms: & $Y^{(1-+)}_1=e^1,$ &$Y^{(1-+)}_2=e^2,$\\
\hline
2-forms: & $Y^{(2--)}_1 =e^{36},$ & $Y^{(2--)}_2 = e^{45}$\\
& $Y^{(2+-)}_1=e^{34},$ & $Y^{(2+-)}_2=e^{56} $\\
& $Y^{(2-+)}_1=e^{35},$ & $Y^{(2-+)}_2=e^{46}$\\
\hline
\end{tabular}
\caption[]{\label{basisforms} One- and two-forms and their transformation properties under the O5- and O7-orientifold projections of table \ref{table:IIBOplanes}.}
\end{center}
\end{table}

\noindent where $e^n,\, n=1,\ldots 6$, are globally defined one-forms on the compact space, and we suppressed the wedge product such that $e^{nm}=e^n \w e^m$. We choose the orientation of the internal manifold such that $1 = -\int e^{123456}$ which then gives
\be\label{eq:intersections}
\tilde{X}_{12} = \tilde{X}_{21} =-1, \quad \hat{X}_{12} = \hat{X}_{21} =-1, \quad \bar{X}_{12} = \bar{X}_{21} = 1,
\ee
with the other components vanishing. The metric fluxes are defined by $\dd e^n =-\frac12 f^n_{mp} e^m \wedge \e^p$ so that we find
\ba\label{eq:rmatrices}
\left( \begin{array}{c} r^1_a \\
                         r^2_a \\
                       \end{array}
                     \right) &=& \left( \begin{array}{c} -f^a_{35}\\
                      -f^a_{46}\\ \end{array} \right),\nn \\
\left( \begin{array}{cc} \tilde{r}^{a1}_1 & \tilde{r}^{a1}_2 \\
                         \tilde{r}^{a2}_1 & \tilde{r}^{a2}_2 \\
                       \end{array}
                     \right) &=& \left(
                       \begin{array}{cc}
                         -f^6_{a4} & f^5_{a3} \\
                         -f^3_{a5} & f^4_{a6} \\
                       \end{array}
                     \right),\\
\left( \begin{array}{cc} \hat{r}^{a1}_1 & \hat{r}^{a1}_2 \\
                         \hat{r}^{a2}_1 & \hat{r}^{a2}_2 \\
                       \end{array}
                     \right)&=& \left(
                       \begin{array}{cc}
                         -f^4_{a6} & -f^5_{a3} \\
                         f^3_{a5} & f^6_{a4} \\
                       \end{array}
                     \right),\nn
\ea
for $a=1,2$.

We now discuss four explicit examples of coset spaces with SU(2)-structure and then discuss examples obtained by twisting $T^2 \times T^4/\mathbb{Z}_2$. For the simplest coset example we solve the F-term equations \eqref{eq:Fterms} explicitly and obtain fully stabilized supersymmetric AdS vacua with large volume and small string coupling.

\subsection{$\mathbf{\frac{SU(3) \times U(1)}{SU(2)}}$\label{SU(3)U(1)}}

For this model, the non-vanishing metric fluxes relevant for our compactification are $f^1_{35}=-f^1_{46}=\sqrt{3}/2$, cyclic. The left-invariant two-forms in the presence of the O5- and O7-planes of table \ref{table:IIBOplanes} read \footnote{We have relabeled the vielbeine used in \cite{Koerber:2008rx,Caviezel:2008tf} so that the two one-forms of the SU(2)-structure are $e^1$ and $e^2$.}:

\bigskip
\begin{center}
  \begin{tabular}{|c|c|c|}
    \hline
    type under O5/O7& basis & name \\
    \hline
    \hline
  odd/even 1-form  &  $ e^{1},  e^{2} $ & $Y^{(1-+)}_a$ \\
 \hline
   odd/odd 2-form  &  $e^{36}+e^{45}$ & $Y^{(2--)}$\\
    \hline
  even/odd 2-form  &  $e^{34}+e^{56}$ & $Y^{(2+-)}$\\
 \hline
 odd/even 2-form &  $ e^{35}-e^{46}$ & $Y^{(2-+)}$ \\
    \hline
   \end{tabular}
\end{center}

This means that, compared with the more generic discussion above, we have only half as many two-forms and hence $t^1=t^2\equiv t$, $z^1=z^2\equiv z$ and $w^1 = -w^2 \equiv w$. Choosing $x>0$ and $k>0$, the necessary condition for metric positivity \footnote{We refer the interested reader to \cite{Grana:2006hr,Grana:2005ny,Grana:2006kf} for the calculation of the metric.}  is $u v < 0$.

Next we expand the background fluxes in our basis. According to \eqref{eq:fluxes} we get for the RR fluxes \footnote{In order to avoid confusion between the one-forms $e^n$ and the expansion forms $e^{ai}$ of the $F_3$ flux, we use a different label in the expansion of the $F_3$-flux. The minus sign in front of $f^1$ is due to $Y^{(4++)} = -e^{3456}$ (c.f. \eqref{eq:fluxes}).}
\eq{\label{FLUXIIBsuusu}\spl{
F_1  &=  m^1 e^{1} + m^2 e^{2} \, ,\\
F_3  &=  f^{(3)} (e^{236} + e^{245}) \, , \\
F_5  &=  -f^1 e^{13456} \, ,
}}
where the exact parts of $F_3$ and $F_5$ (i.e. the $(e^{136}+e^{145})$ part of $F_3$ and the $e^{23456}$ part of $F_5$) are put to zero since they can be absorbed into a shift of $C_2$  and $C_4$, respectively. For the $H_3$-flux we choose
\eq{
H_3  =  0 \, , \\
}
since the absence of NS5-branes requires the closure of $H_3$ and the exact part can be absorbed into a shift of $B$ (the three-forms have a trivial cohomology as $b_3=0$ \cite{Caviezel:2008ik}). Using the expression for the superpotential \eqref{eq:superpotential2} we calculate
\eq{
W = -\frac{i}{2} \lp -2 f^{(3)} t + i m^2 {(t)}^2 + \sqrt{3} w + f^1 \tau - 2 \sqrt{3} t \, z \, \tau + m^1 {(t)}^2 \tau \rp \,.
}
For the K\"ahler potential we obtain from \eqref{eq:Kaehlerpotential} and \eqref{eq:Kaehlerpotential2}
\eq{\spl{
K  = &  -\ln \left((\tau+\bar{\tau}) (t+\bar{t})^2 \right)
- \ln \left(\frac{1}{16} ( z+\bar{z})^2 ( w+\bar{w})^2 \right)\, .
}}
To demonstrate that it is possible to stabilize all moduli in a supersymmetric AdS vacuum, we will explicitly solve the F-term equations \eqref{eq:Fterms} for this very simple model. From the K\"ahler and superpotential given above we find
\ba
0 &=& D_\tau W =-\frac{i}{2} \lp f^1 - 2 \sqrt{3} t \, z + m^1 {(t)}^2 \rp- \frac{1}{2x}W, \\
0 &=& D_{t} W = i f^{(3)}+ m^2 t +i \sqrt{3} z \, \tau -i m^1 t\, \tau - \frac{1}{k}W,\\
0 &=& D_{z} W = i \sqrt{3} t \, \tau - \frac{1}{u}W,\\
0 &=& D_{w} W = -i\frac{\sqrt{3}}{2} - \frac{1}{v}W.
\ea
From $D_{w} W =0$ we find Re$(W) = 0$ which gives
\ba
\sqrt{3} c_{(4)} &=& - 2 f^{(3)} b- f^1 y+ b^2 (m^2+ m^1 y)+ 2 b \left(k m^1 x+ \sqrt{3} (-u x+ c_{(2)} y)\right) \nn \\
&&-k \left(k (m^2+m^1 y)-2 \sqrt{3} (c_{(2)} x + u y)\right).
\ea
Since Re$(W)=0$ we can easily solve Re$(D_z W)= - \sqrt{3} (k\, y-b\,x) = 0$ to find
\be
b=\frac{k y}{x},
\ee
Re$(D_{t} W)=0$ to find
\be
k=\frac{\sqrt{3} (c_{(2)} x +u\, y)}{m^2},
\ee
and Re$(D_{\tau} W)=0$ which leads to
\be
c_{(2)}=\frac{u y (m^2-m^1 y)}{x (m^2+m^1 y)}.
\ee
Next we solve Im$(D_{t} W)=0$ and find
\be
v=\frac{-f^1 x^2 (m^2+m^1 y)^2+12 u^2 y^2 \left(2 m^2 y+m^1 \left(x^2+y^2\right)\right)}{\sqrt{3} x (m^2+m^1 y)^2}
\ee
and from Im$(D_{z} W)=0$ we get
\be
u= \frac{f^{(3)} x (m^2+m^1 y)}{2 \sqrt{3} y \left(2 m^2 y+m^1 \left(x^2+y^2\right)\right)}.
\ee
Finally, we solve Im$(D_{\tau} W)=0$, which is a quartic polynomial in $x$. There are two positive solutions, and we choose
\ba
x&=&\frac{1}{2\sqrt{f^1 \lp m^1 \rp^2 y}} \Bigg[ \lp f^{(3)} \rp^2 (m^2-m^1 y)-4 f^1 m^1 y^2 (2 m^2 +m^1 y)  \nn \\
&&  -f^{(3)} \sqrt{m^2-m^1 y} \sqrt{-16 f^1 m^1 m^2 y^2+ \lp f^{(3)} \rp^2 (m^2-m^1 y)} \Bigg]^{1/2}.
\ea
We are now left with
\ba
0&&=\text{Im}(D_w W) \nn\\
&&=\frac{\sqrt{3} \left(-f^1 y \left(2 m^2 y+m^1 \left(x^2+y^2\right)\right)^2+\lp f^{(3)} \rp^2 \left(2 m^1 y \left(x^2+y^2\right)+m^2 \left(x^2+3 y^2\right)\right)\right)}{2y \left(2 m^2 y+m^1 \left(x^2+y^2\right)\right) \left(-\lp f^{(3)} \rp^2+f^1 \left(2 m^2 y+m^1 \left(x^2+y^2\right)\right)\right)}, \nn
\ea
where $x$ is as given above. One can solve this equation analytically for $y$. However, the resulting expression is not very illuminating and rather long so we do not write it down explicitly.\\
The fluxes in the solution above are constrained by the tadpole conditions \eqref{eq:tadpole}
\ba
\dd F_1 &=& - \frac{\sqrt{3}}{2} m^1 (e^{35}-e^{46}) = 4 [\delta_{O7}] = 4 N_{O7} (e^{35}-e^{46}), \nn \\
\dd F_3 + H_3 \w F_1 &=& \sqrt{3} f^{(3)}(e^{1234}+e^{1256}) = [\delta_{O5}] = N_{O5} (e^{1234}+e^{1256}), \nn
\ea
where we have expanded the orientifold contributions in our forms. So we see that the tadpole condition fixes $m^1$ and $f^{(3)}$. However, the fluxes $m^2$ and $f^1$ are still unconstrained and one can make them large. Solving $\text{Im}(D_w W)=0$ as given above gives a scaling of $y \sim |f^1|^0 |m^2|^1$ for large $|f^1|$ and $|m^2|$. From this and the solutions for the other moduli we find
\ba
x &\sim& |f^1|^0 |m^2|^1, \nn \\
u &\sim& |f^1|^{1/2} |m^2|^0, \nn \\
v &\sim& |f^1|^1 |m^2|^1, \nn \\
k &\sim& |f^1|^{1/2} |m^2|^0. \nn
\ea
This leads to the following scaling for the four- and ten-dimensional dilaton
\ba
e^{-D} &\sim& \sqrt{|u v|} \sim |f^1|^{3/4} |m^2|^{1/2},\nn\\
e^{-\phi} &=& \frac{e^{-D}}{\sqrt{vol_6}} \sim \sqrt{\frac{|uv|}{xk^2}} \sim |f^1|^{1/4} |m^2|^0.\nn
\ea
So we see that if we make simultaneously $|f^1|$ and $|m^2|$ large then we will have a large volume with $x, k \gg 1$ and small four- and ten-dimensional dilaton $e^{D}, e^{\phi} \ll 1$. In this limit we can trust our supergravity analysis and we have therefore found a large number of trustworthy, fully stabilized supersymmetric AdS vacua. We present a few more explicit models but leave it to the interested reader to solve the F-term equations and find fully stabilized AdS vacua for the other models.

\subsection{$\mathbf{\frac{SU(2)^2}{U(1)} \times U(1)}$}\label{SU(2)2U(1)}

The non-vanishing metric fluxes relevant for this compactification are $f^1_{35}=1$, cyclic, and $f^1_{46}=-1$. The left-invariant two-forms in the presence of our O5- and O7-planes are

\bigskip
\begin{center}
  \begin{tabular}{|c|c|c|}
    \hline
    type under O5/O7& basis & name \\
    \hline
    \hline
  odd/even 1-form  &  $ e^{1},  e^{2} $ & $Y^{(1-+)}_i$ \\
 \hline
 odd/odd 2-form  &  $e^{36}+e^{45}$ & $Y^{(2--)}$\\
 \hline
 even/odd 2-form  &  $e^{34}+e^{56}$ & $Y^{(2+-)}$\\
 \hline
 odd/even 2-form &  $ e^{35}$ , $e^{46}$ & $Y^{(2-+)}_I$ \\
    \hline
   \end{tabular}
\end{center}

This means that comparing with the more generic model above we have to set $t^1=t^2 \equiv t$ and $z^1=z^2\equiv z$. For the background fluxes we get
\eq{\spl{
H_3  &=  0 \, , \\
F_1  &=  m^1 e^{1} + m^2 e^{2} \, ,\\
F_3  &=  f^{(3)} (e^{236} + e^{245}) \, , \\
F_5  &=  f^1 e^{13456} \, ,
}}
where we again set the exact parts of $H_3$, $F_3$ and $F_5$ to zero and chose $H_3$ to be closed. The superpotential thus reads
\eq{
W =-\frac{i}{2} \lp - 2 f^{(3)} t + i m^2 (t)^2 + w^1 - w^2 + f^1 \tau - 2 t \,z \,\tau + m^1 (t)^2 \tau \rp\,  .
}
For the K\"ahler potential we obtain
\eq{
K  =   -\ln \left((\tau+\bar{\tau}) (t+\bar{t})^2 \right)
- \ln \left(-\frac{1}{16} (z+\bar{z})^2 (w^1+\bar{w}^1)(w^2+\bar{w}^2) \right)\, .
}
Necessary conditions for metric positivity are $x>0$, $k>0$ and $v^1 v^2 < 0$, $u v^1 <0$.

\subsection{$\mathbf{SU(2) \times SU(2)}$}\label{SU(2)SU(2)}

The non-vanishing metric fluxes relevant for the compactification on SU(2)$\times$SU(2) are $f^1_{35} = f^2_{46}=1$,  cyclic. For this model, the forms given in table \ref{basisforms} form the basis for the left-invariant one- and two-forms. The background fluxes are chosen
\eq{\spl{
H_3  &=  0 \, , \\
F_1  &=  m^1 e^{1} + m^2 e^{2} \, ,\\
F_3  &=  f^{(3)1} (e^{136} + e^{245})+f^{(3)2} (e^{145} +  e^{236}) \, , \\
F_5  &=  0 \, .
}}
The superpotential reads for this choice
\eq{\spl{
W &= -\frac{i}{2} \Big( f^{(3)1} (-t^1 + i t^2 \tau) + f^{(3)2} (-t^2 + i t^1 \tau) +i m^2 t^1 t^2 + m^1 t^1 t^2 \tau \\
& + i t^1 z^2 + i t^2 z^1 - t^1 z^1 \tau - t^2 z^2 \tau - i w^1 \tau - w^2\Big)\,  .
}}
The K\"ahler potential reads
\eq{\spl{\label{KaIIBsusu}
K &=  -\ln \left((\tau+\bar{\tau}) (t^1+\bar{t}^1)(t^2+\bar{t}^2) \right)
- \ln \left(-\frac{1}{16} (z^1+\bar{z}^1)(z^2+\bar{z}^2) ( w^1+\bar{w}^1)(w^2+\bar{w}^2) \right).
}}
Necessary conditions for metric positivity are $x>0$, $k^1>0$, $k^2>0$ and $v^1 v^2 < 0$, $u^1 u^2 >0$ and $u^2 v^2 > 0$.

\subsection{$\mathbf{SU(2) \times U(1)^3}$}\label{SU(2)U(1)3}

The analysis of this model is quite similar to the analysis of the model SU(2)$\times$SU(2), as one only turns off the structure constant $f^{2}_{46} = 0$. Therefore, we choose the same expansion forms as in the model above. The only difference is in the choice of background fluxes, since the cohomology changes, and we choose
\eq{\spl{
H_3  &=  0 \, , \\
F_1  &=  m^1 e^{1} + m^2 e^{2} \, ,\\
F_3  &=  f^{(3)1} e^{236} + f^{(3)2} e^{245} \, , \\
F_5  &=  f^1 e^{13456}\, ,
}}
such that the superpotential reads
\eq{
W = -\frac{i}{2} \lp - f^{(3)1} t^2 - f^{(3)2} t^1 + i m^2 t^1 t^2 - w^2 + f^1 \tau - t^1 z^1 \tau - t^2 z^2 \tau + m^1 t^1 t^2 \tau \rp\,  ,
}
and the K\"ahler potential is as in eq.~\eqref{KaIIBsusu}.

\subsection{Twisted tori from the base-fiber splitting}\label{twistedtori}
In this subsection we study models that are twisted $T^2 \times T^4/\mathbb{Z}_2$ and can be obtained from a base-fiber construction \cite{Hull:2004in,Dabholkar:2005ve,Grana:2006kf,Hull:2006qs,Hull:2006va,Cvetic:2007ju,Ihl:2007ah,Bergman:2007qq}. The basic idea for this construction is to split our compact space into a base space and a fiber. If we know the T-duality group of the fiber, we can associate a T-duality element to every non-trivial cycle in the base space. Then we can twist the fiber by this T-duality element, if we go around the non-trivial cycle in the base. The resulting space is often called ``twisted torus", if one starts with a torus (or, as in our case, a toroidal orientifold), but the reader should keep in mind that the new space has in general nothing to do with a torus anymore. It is a different topological space. For a twist by a generic T-duality element the resulting space is only locally geometric. Since the underlying string theory is invariant under T-duality one expects that the new ``space" is nevertheless still a good string background. We however are working on the level of supergravity and need a geometric compactification space for our analysis to be valid. Therefore, we restrict ourselves to twists by T-duality elements that are generated by the geometric subgroup of the T-duality group (for a torus GL$(d;\mathbb{Z}) \subset$ SO$(d,d;\mathbb{Z})$) and shifts of the $B$ field. Although, the basic idea is very simple the explicit calculations are fairly lengthy. We therefore present only the results in the main body of the paper and work out the details of one explicit example in appendix \ref{app:basefiber}.\\
To do a full classification one needs to consider all possible splittings of the $T^2 \times T^4/\mathbb{Z}_2$ into a base space and a fiber. This splitting needs to be invariant under the orbifold and orientifold projection. We consider all cases in which the basis and fiber are parallel or perpendicular to the coordinate axis. Due to the symmetry of our space there are 19 different possibilities. Denoting for example a one-dimensional base extending along the first direction by $\{1\}$ they are $\{1\}$, $\{3\}$, $\{1,2\}$, $\{1,3\}$, $\{3,4\}$, $\{3,5\}$, $\{3,6\}$, $\{1,2,3\}$, $\{1,3,4\}$, $\{1,3,5\}$, $\{1,3,6\}$, $\{3,4,5\}$, $\{1,2,3,4\}$, $\{1,2,3,5\}$, $\{1,2,3,6\}$, $\{1,3,4,5\}$, $\{3,4,5,6\}$, $\{1,2,3,4,5\}$, $\{1,3,4,5,6\}$. Calculating the possible metric fluxes resulting from twisting the fiber over these base spaces we find the following non-vanishing NSNS fluxes:\\
\allowdisplaybreaks\ba\label{eq:NSNSfluxes}
&\{1\}: & H_{134},\, H_{156},\, f^3_{15},\,  f^4_{16},\, f^5_{13},\, f^6_{14} \nn \\
&\{3\}: & H_{134},\, H_{234},\, f^1_{35},\,  f^2_{35},\, f^5_{13},\, f^5_{23} \nn \\
&\{1,2\}: & H_{134},\, H_{156},\, H_{234},\, H_{256},\, f^3_{15},\, f^3_{25},\, f^4_{16},\, f^4_{26},\, f^5_{13},\, f^5_{23},\,f^6_{14},\,f^6_{24}\nn \\
&& f^3_{25} f^5_{13}-f^3_{15} f^5_{23} =0, \,\, f^4_{26} f^6_{14} -f^4_{16} f^6_{24}=0\nn \\
&\{1,3\}: & H_{156},\, H_{234},\, f^2_{35},\,  f^4_{16},\, f^5_{23},\, f^6_{14}\nn \\
&\{3,4\}: & f^1_{35},\,  f^1_{46},\, f^2_{35},\, f^2_{46},\, f^5_{13},\,  f^5_{23},\, f^6_{14},\,  f^6_{24},\,\nn \\
&& f^1_{35} f^6_{14}+f^2_{35} f^6_{24} =0, \,\, f^1_{46} f^5_{13} +f^2_{46} f^5_{23}=0\nn \\
&\{3,5\}: & H_{134},\, H_{156},\, H_{234},\, H_{256}\nn \\
&\{3,6\}: & H_{134},\, H_{156},\, H_{234},\, H_{256},\, f^1_{35},\, f^1_{46},\, f^2_{35},\, f^2_{46},\, f^4_{16},\, f^4_{26},\,f^5_{13},\,f^5_{23}\nn \\
&& f^1_{35} f^4_{16}+f^2_{35} f^4_{26} =0, \,\, f^1_{46} f^5_{13} + f^2_{46} f^5_{23}=0\nn \\
&\{1,2,3\}: & H_{156},\, H_{256},\, f^4_{16},\,  f^4_{26},\, f^6_{14},\, f^6_{24}\nn \\
&& f^4_{16} f^6_{24}-f^4_{26} f^6_{14} =0\nn \\
&\{1,3,4\}: & H_{156},\, f^2_{35},\,  f^2_{46},\, f^5_{23},\, f^6_{24}\nn \\
&& f^2_{35} f^6_{24}=0, \,\, f^2_{46} f^5_{23} =0\nn \\
&\{1,3,5\}: & H_{234},\, H_{256},\, f^4_{16},\,  f^6_{14}\nn \\
&\{1,3,6\}: & H_{234},\, H_{256},\, f^2_{35},\,  f^2_{46},\, f^4_{26},\, f^5_{23}\nn \\
&& f^2_{35} f^4_{26}=0, \,\, f^2_{46} f^5_{23} =0\nn \\
&\{3,4,5\}: & H_{156},\, H_{256},\, f^1_{46},\,  f^2_{46},\, f^6_{14},\, f^6_{24}\nn \\
&\{1,2,3,4\}: & H_{156},\, H_{256}\nn \\
&\{1,2,3,5\}: & f^4_{16},\,  f^4_{26},\, f^6_{14},\, f^6_{24}\nn \\
&& f^4_{26} f^6_{14}-f^4_{16} f^6_{24} =0\nn \\
&\{1,3,4,5\}: & H_{256},\, f^2_{46},\,  f^6_{24}\nn \\
&\{1,2,3,6\}, &\{3,4,5,6\}, \, \{1,2,3,4,5\}, \, \{1,3,4,5,6\}:\quad \text{No } H_3 \textrm{- or metric flux}\nn
\ea
where we also spelled out any constraints arising from the Bianchi identities \eqref{eq:Bianchi}.

The explicit expansion forms for these models are given in table \ref{basisforms}. The superpotential for these models is as given in \eqref{eq:superpotential2} and the K\"ahler potential is given in \eqref{eq:Kaehlerpotential}, \eqref{eq:Kaehlerpotential2} where all indices run from 1 to 2 and the intersection numbers are given in \eqref{eq:intersections}. The NSNS fluxes from above lead to $h^{11}=H_{134}$, $h^{12}=H_{156}$, $h^{21}=H_{234}$, $h^{22}=H_{256}$ and the metric flux matrices are as given in \eqref{eq:rmatrices}. The necessary conditions for metric positivity are $x>0$, $k^1>0$, $k^2>0$ and $v^1 v^2 < 0$, $u^1 u^2 >0$ and $u^2 v^2 > 0$.

\section{Cosmological aspects of SU(2)-structure compactifications of type IIB supergravity}\label{nogos}

As we have seen in the previous sections, type IIB compactifications on SU(2)-structure manifolds have the interesting property that their scalar potentials generically depend on all moduli at tree level and allow for stabilized supersymmetric anti-de Sitter vacua. In this section we discuss whether these potentials could also provide cosmologically interesting solutions such as meta-stable de Sitter vacua and/or slow-roll inflation models in some other regions of moduli space. We start out by showing that our general setup of SU(2)-structure manifolds with O5- and O7-planes evades previous general no-go theorems against dS vacua and slow-roll inflation \cite{Hertzberg:2007wc,Haque:2008jz,Danielsson:2009ff} in tree-level IIB compactifications. For models that are formally T-dual to type IIA compactifications on SU(3)-structure manifolds with O6-planes and \emph{no} non-geometric fluxes the no-go theorems derived in \cite{Hertzberg:2007wc,Haque:2008jz,Caviezel:2008tf,Flauger:2008ad,Danielsson:2009ff} on the IIA side could be used to rule out also the corresponding IIB models where applicable. We will derive a number of no-go theorems directly in type IIB, some of which can be viewed as the IIB translations of the above mentioned IIA theorems. In general, however, the IIB compactifications studied here are formally T-dual to \emph{non}-geometric type IIA compactifications, and hence the corresponding no-go theorems proven here extend the IIA no-go theorems of \cite{Hertzberg:2007wc,Caviezel:2008tf,Flauger:2008ad}. The dS vacua of \cite{deCarlos:2009qm} with non-geometric fluxes together with the unstable dS extrema of \cite{Caviezel:2008tf,Flauger:2008ad} suggest that there is no general no-go theorem for the class of compactifications we are considering in this paper.

\subsection{Review of previous no-go theorems in type IIB}

We start by reviewing the scaling of the terms in type IIA/IIB supergravity with respect to $\rho \equiv (vol_6)^{1/3}$ and $q \equiv e^{-D}= e^{-\phi} \sqrt{vol_6} = e^{-\phi} \rho^{3/2}$ on which the no-go theorems of \cite{Hertzberg:2007wc} are based. The Einstein term in string frame is
\be
S \supset \frac{1}{2} \int d^{10}x \sqrt{-g^{(10)}} e^{-2 \phi}R^{(10)} = \frac{1}{2} \int d^{4}x \sqrt{-g^{(4)}} vol_6 e^{-2 \phi} \left( R^{(4)} + \ldots \right),
\ee
so that we have to define $g^{(4)}_{\mu \nu} = \frac{1}{q^2} g^{(E)}_{\mu \nu}$ to go to the four-dimensional Einstein frame
\ba
S \supset \frac{1}{2} \int d^{4}x \sqrt{-g^{(4)}} vol_6 e^{-2 \phi} \left( R^{(4)} + \ldots \right)= \frac{1}{2} \int d^{4}x \sqrt{-g^{(E)}} \left( R^{(E)} + \ldots \right).
\ea
 From the type II supergravity action
\be
S = \frac{1}{2} \int d^{10}x \sqrt{-g^{(10)}} \left(e^{-2 \phi} \left(R^{(10)} + 4 (\partial_\mu \phi)^2 - \frac12 |H_3|^2 \right) - \sum_p |F_p|^2\right)
\ee
and the scaling $g^{(6)}_{\mu \nu} \sim \rho$, we then find the following scaling for the contributions to the four-dimensional scalar potential coming from the $H_3$ and RR fluxes as well as the Ricci scalar,
\ba
V_H &\sim& q^{-2} \rho^{-3}, \nn\\
V_p &\sim& q^{-4} \rho^{3-p}, \\
V_f &\sim& q^{-2} \rho^{-1}.\nn
\ea
Note that $V_H, V_p \geq 0$ while $V_f$ can have either sign.

For Dp-branes and Op-planes we find from $S_{Dp/Op} \sim \pm \mu_p \int d^{p+1}x \sqrt{-g} e^{-\phi}$ that
\be
V_{Dp/Op} \sim \pm q^{-3} \rho^{\frac{p-6}{2}},
\ee
where $V_{Dp} \geq 0$ and $V_{Op} \leq 0$. Although we will not use it here, we also summarize the scaling for NS5-branes and KK-monopoles, which can be obtained from $S_{NS5} \sim \int d^6x \sqrt{-g} e^{-2 \phi}$ and $S_{KK} \sim \int d^7x \sqrt{-g} e^{-2 \phi} g^{\psi \psi}$ \cite{Hertzberg:2007wc,Villadoro:2007tb}
\ba
V_{NS5} &\sim& q^{-2} \rho^{-2}, \nn \\
V_{KK} &\sim& q^{-2} \rho^{-1}.
\ea
In principle one could also consider the addition of so called non-geometric $Q$- and $R$-flux \cite{deCarlos:2009fq,deCarlos:2009qm} (or even the corresponding sources). By T-duality arguments one finds
\ba
V_Q &\sim& q^{-2} \rho^1, \nn \\
V_R &\sim& q^{-2} \rho^3.
\ea
For type IIB compactifications with only D3/O3-sources, $H_3$- and $F_3$-flux, all terms scale like $\rho^{-6}$ times a certain power of the dilaton $e^\phi$ (not $q$), and  one finds the simple relation \cite{Hertzberg:2007wc}
\be
- \rho \partial_\rho V = 6V,
\ee
which shows that de Sitter vacua cannot exist. One can however choose the dilaton and complex structure moduli such that the potential $V$ vanishes, corresponding to Minkowski vacua with $\rho$ being a classically flat direction. These are of course nothing but the no-scale solutions of \cite{Giddings:2001yu}, and quantum effects as in \cite{Kachru:2003aw} may be used to change this behaviour. If one wants to stay at a purely classical level, this simple argument against de Sitter vacua or inflation could in principle also be circumvented by including metric fluxes (or O$p$-planes for $p\neq3$).

A natural question therefore is whether one can obtain cosmologically interesting models by studying type IIB on SU(3)-structure manifolds (first assuming only O3-planes). Due to the extra terms involving the metric fluxes \cite{Grana:2005jc,Benmachiche:2006df,Robbins:2007yv} it might be possible to stabilize the geometric moduli. Furthermore, using the scalings given above we can show that it is impossible to have a no-go theorem in the $(\rho,q)$-plane for type IIB on SU(3)-structure manifolds with O3-planes, $H_3$- and $F_3$-fluxes. Indeed, for a no-go theorem we would have to show that, along an arbitrary direction in the $(\rho,q)$-plane, we have
\be
DV\equiv(a \p_\rho + b \p_q) V \geq c \, V, \quad c>0.
\ee
Since $V_H, V_3 \geq 0$, $V_{O3} < 0$ and $V_f$ can have either sign we need to find $a,b$ such that
\be
D V_{H/3} = c_{H/3} V_{H/3}, \quad D V_{O3} = c_{O3} V_{O3}, \quad D V_f = c V_f, \quad c_{H},c_{3} \geq c, \quad c_{O3} \leq c.
\ee
It is straightforward to check that there is no such solution. So in principle these models seem interesting and deserve further study. However, in our concrete examples we restrict ourselves to the bulk moduli for simplicity. If one wants to have a bulk O3-plane then one needs an orientifold projection that reverses all six coordinates. Since the metric fluxes $f^p_{mn}$ have to be invariant under the orientifold projection, one finds that bulk O3-planes and bulk metric fluxes are incompatible. Therefore, for interesting examples in this direction one would have to include for example blow-up modes and  have either the O3-planes or the metric fluxes being nontrivial in the twisted sector only \footnote{We thank D. Robbins and G. Dall'Agata for discussions on this point.}.\\
In \cite{Haque:2008jz} the authors look at the quantity $\frac{4 ac}{b^2}$ in type IIA compactifications, where $a$ contains all terms that scale with $q^{-2}$, $b$ all terms that scale like $q^{-3}$ and $c$ all terms that scale like $q^{-4}$. This quantity is therefore independent of $q$, and one can show that a de Sitter minimum requires that $\frac{4 ac}{b^2}$ has a minimum in $\rho$ and the remaining moduli at which $\frac{4 ac}{b^2}\approx 1$ \cite{Silverstein:2007ac}. In type IIA without non-geometric fluxes and only O6/D6 sources one finds
\be
\frac{4 a c}{b^2} \sim \sum_p A_p \left(A_H \rho^{-p}+A_f \rho^{2-p} \right),
\ee
where the $A$s are coefficients such that for example $V_p = A_p q^{-4} \rho^{3-p}$. From the positivity of $V_H$ we have that $A_H\geq0$. To evade a no-go theorem from \cite{Hertzberg:2007wc,Silverstein:2007ac,Caviezel:2008ik} one would need $A_f>0$ as well, which corresponds to spaces with negative scalar curvature (see also \cite{Neupane:2005nb}). If there is no flux with $p<2$, then the minimum in the $\rho$ direction is at $\rho = \infty$, i.e. we have a runaway direction that leads to a decompactification. So the conclusion is that, in the above setup, one needs the Romans mass parameter in type IIA.

In \cite{Haque:2008jz} it is further speculated that, by analogy to the Romans mass for type IIA, the $F_1$ flux might be useful for classical type IIB de Sitter vacua.
So let us look at type IIB compactifications with $H_3$-flux, $F_p$-flux, metric flux and only O$n$ sources for one fixed $n$. We find
\be
\frac{4 a c}{b^2} \sim \sum_p A_p \left(A_H \rho^{-p} + A_f \rho^{2-p}\right) \rho^{6-n}.
\ee
So we see that one needs $F_1$ flux for $n = 5$, while for the standard case with O3-planes $F_3$ flux is sufficient to avoid this no-go theorem, as we have already shown above. For cases with only $n =7$ or $n=9$ O-plane sources, not even $F_1$ flux is sufficient to avoid the no-go theorem.

So, to summarize, we have argued that type IIB compactifications on SU(3)-structure manifolds with O3-planes might in principle be cosmologically interesting, but they would require to go beyond the study of just bulk moduli. Using instead O5-planes, one would need $F_1$-flux, which, however, cannot be turned on on an SU(3)-structure manifold unless it is actually an SU(2)-structure manifold. As we have also shown that having only one kind of O$p$-plane with $p=7,9$ would not work, it follows that IIB compactifications on strict SU(3)-structure manifolds cannot lead to classical dS vacua or slow-roll inflation if we have only one kind of O$p$-planes with $p =5,7,9$, or they would require going beyond the bulk moduli for $p=3$.

Turning to cases with two different types of O$p$-planes (and requiring some unbroken supersymmetry in the action), the only other potentially interesting SU(3)-structure compactifications then have either O3- and O7-planes, or O5- and O9-planes. For the O3/O7-plane case one has again the problem we discussed above due to the O3-planes. The case with O5- and O9-planes is beyond the scope of this paper, as the O9-plane charge cannot be canceled by fluxes, and we therefore have to introduce D9-branes and worry about open string moduli. So our SU(2)-structure compactifications with $F_1$-flux and O5/O7-planes remain as a very tractable class of models that evades all previously discussed no-go theorems.

\subsection{No-go theorems for type IIB compactifications on SU(2)-structure manifolds}\label{nogo}
In this section we will derive several no-go theorems for type IIB compactifications of type IIB on SU(2)-structure manifolds in the presence of O5- and O7-planes. The idea behind the no-go theorems is to find a direction in moduli space along which the slope is of the same order as the value of the potential which then leads to a slow-roll parameter $\e$ of order one. This then excludes slow-roll inflation as well as de Sitter vacua.

The directions in moduli space we look at will generically involve the dilaton and some of the K\"ahler moduli. We also use the complex structure
moduli but never the axionic moduli coming from $C_2, C_4$ and $B$. The no-go theorems will be in the spirit of the no-go theorems derived in
\cite{Flauger:2008ad,Caviezel:2008tf} for type IIA compactifications on SU(3)-structure manifolds. There the authors used a split of the
K\"ahler and/or complex structure moduli into two sets. A priori such a split seems rather unnatural for a generic SU(3)-structure manifold, but among the cosets spaces and twisted tori spaces studied in \cite{Flauger:2008ad,Caviezel:2008tf} this split turns out to be generically present. In the SU(2)-structure manifolds we study here, the split likewise appears naturally. The K\"ahler moduli split into $\tau$ and the $t^i$, while the complex structure moduli split into the $z^A$ and the $w^I$. We will make use of this split by studying directions in moduli space that only involve for example $\tau$ and not the $t^i$. In the concrete models we will study there are further splits since the intersection forms $\tilde{X}_{ij}, \hat{X}_{AB}, \bar{X}_{IJ}$ have only off diagonal entries so that for example $\tilde{X}_{ij} t^i t^j =-2 t^1 t^2$ is compatible with a further split of the $t^i$ into $t^1$ and $t^2$. This allows us to study also directions that involve for example only $t^1$ but not $t^2$.

Generically the no-go theorems we derive will only apply to models that satisfy some restrictions on the NSNS fluxes. Such restrictions set in certain cases the flux contribution to the O5- and/or O7-tadpole conditions to zero. This requires one to introduce D-branes or consider cases with fewer
O-planes that potentially preserve more supersymmetry in four dimensions. This is beyond the scope of this paper. Nevertheless, we write down these
no-go theorems since they make use of the natural splitting in the K\"ahler and complex structure sector and they apply to some of the twisted tori one can obtain from the base-fiber splitting.\\
To start with, we calculate the explicit scalar potential \eqref{eq:scalarpotential} using the K\"ahler potential \eqref{eq:Kaehlerpotential},
\eqref{eq:Kaehlerpotential2} and superpotential \eqref{eq:superpotential}. We redefine
\begin{equation}
u^A = q \,s\, \mathcal{U}^A, \quad v^I =\frac{q}{s} \mathcal{V}^I,
\end{equation}
with $q=e^{-D}$ such that $\hat{X}_{AB} \mathcal{U}^A \mathcal{U}^B \bar{X}_{IJ} \mathcal{V}^I \mathcal{V}^J =4$. We have summarized a couple of
useful results related to the K\"ahler potential in appendix \ref{app:Kaehler}. Using these one finds after a long but straightforward calculation the following contributions to the scalar potential
\ba\label{eq:scalarpotential2}
V_H &=& \frac{s^2}{16 (-\tilde{X}_{ij} k^i k^j) \, q^2} \hat{K}^{AB} \hat{X}_{AC} \hat{X}_{BD} M_{ab} \lp h^{aC} - \tilde{r}^{aC}_i b^i \rp \lp h^{bD}
- \tilde{r}^{bD}_i b^i \rp, \\
V_f &=& \frac{1}{16 (-\tilde{X}_{ij} k^i k^j) \, q^2} \Bigg\{ M_{ab} \Bigg[- \frac{1}{s^2} \e^{ac} \e^{bd} r_c^I r_d^J \bar{X}_{IK} \bar{X}_{JL}
\bar{K}^{KL} + \frac{s^2}{4} \tilde{r}_i^{aA} \tilde{r}_j^{bB} \tilde{K}^{ij} \hat{X}_{AC} \hat{X}_{BD} \hat{K}^{CD} \nn \\
&& - s^2 \hat{X}_{AB} (\tilde{X}^{-1})^{ij} \tilde{r}^{aA}_i \tilde{r}^{bB}_j \tilde{X}_{kl} k^k k^l \hat{X}_{CD} \mathcal{U}^C \mathcal{U}^D \Bigg]
-8 \bar{X}_{IJ} \tilde{X}_{ij} k^i r_a^I \hat{r}^{aj}_A \mathcal{U}^A \mathcal{V}^J \Bigg\}, \\
V_1 &=& \frac{(-\tilde{X}_{ij} k^i k^j) M_{ab} m^a m^b}{16 q^4}, \\
V_3 &=& \frac{\tilde{K}^{ij} \tilde{X}_{ik} \tilde{X}_{jl} M_{ab} \lp e^{ak} +\hat{r}^{ak}_A c_{(2)}^A- m^a b^k\rp \lp e^{bl} +\hat{r}^{bl}_A
c_{(2)}^A - m^b b^l\rp}{16 (-\tilde{X}_{ij} k^i k^j) \, q^4}, \\
V_5 &=& \frac{1}{4(-\tilde{X}_{ij} k^i k^j) \, q^4} M_{ab} \\
&& \quad \times \lp f^a - \e^{ac} \bar{X}_{IJ} r^I_c c_{(4)}^J + \hat{X}_{AB} h^{aA} c_{(2)}^B - \tilde{X}_{ij} \lp e^{ai} + \hat{r}^{ai}_A c_{(2)}^A
\rp b^j + \hlf m^a \tilde{X}_{ij} b^i b^j \rp \nn \\
&& \quad \times \lp f^b - \e^{bd} \bar{X}_{IJ} r^I_d c_{(4)}^J + \hat{X}_{AB} h^{bA} c_{(2)}^B - \tilde{X}_{ij} \lp e^{bi} + \hat{r}^{bi}_A c_{(2)}^A
\rp b^j + \hlf m^b \tilde{X}_{ij} b^i b^j \rp, \nn\\
V_{O5} &=& -\frac{s \, \e_{ab} \hat{X}_{AB} \mathcal{U}^A \lp e^{ai} \tilde{r}^{bB}_i+ m^a h^{bB} \rp}{4q^3}, \\
V_{O7} &=& -\frac{\bar{X}_{IJ} m^a r^I_a \mathcal{V}^J}{4s\,q^3},\label{eq:scalarpotential3}
\ea
where we used
\ba\label{eq:tau}
(M_{ab})&=& \frac{1}{\text{Re}(\tau)} \left(
    \begin{array}{cc}
      |\tau|^2 & \text{Im}(\tau) \\
      \text{Im}(\tau) & 1 \\
    \end{array}
  \right), \nn\\
\tilde{K}^{ij}&=& K^{i\bar{j}}, \\
\hat{K}^{AB}&=& \frac{1}{s^2 q^2} K^{A\bar{B}}, \nn \\
\bar{K}^{IJ}&=& \frac{s^2}{q^2} K^{I\bar{J}}. \nn
\ea
Here, $K^{*\bar{*}}$ denotes the inverse K\"ahler metric that splits into the four pieces $K^{\tau \bar{\tau}}$, $K^{i \bar{j}}$, $K^{A \bar{B}}$ and
$K^{I \bar{J}}$ as can be seen from the K\"ahler potential \eqref{eq:Kaehlerpotential}, \eqref{eq:Kaehlerpotential2}, and we defined $\hat{K}^{AB}$ and
$\bar{K}^{IJ}$ so that the dependence on $q$ and $s$ is fully explicit in the scalar potential. For $V_{O5}$ we used the Bianchi identities
\eqref{eq:dsquaredIIB} arising from demanding that $\dd^2$ on the forms gives zero, and for $V_H$ we also used that $\dd H_3=0$ (c.f. \eqref{eq:dH}).\\
Note that $\tau = x+iy$ appears in the scalar potential only through $M_{ab}$ and that
\be
M_{ab} A^a A^b = \frac{1}{x} \ls x^2 (A^1)^2 + (A^2 + A^1 y)^2 \rs \geq 0,
\ee
so that $V_H, V_1, V_3, V_5 \geq 0$, since metric positivity requires $x>0$ and in our conventions $-\tilde{X}_{ij} k^i k^j>0$. \\
The expression for $\e$ is
\be
\e = V^{-2}\left\{ K^{\tau \bar \tau}\frac{\p V}{\p \tau} \frac{\p V}{\p\bar{\tau}} + K^{i \bar j}\frac{\p V}{\p t^i}\frac{\p V}{\p\bar{t}^{\bar j}} +
K^{A\bar B}\frac{\p V}{\p z^A} \frac{\p V}{\p\bar{z}^{\bar B}} + K^{I\bar J}\frac{\p V}{\p w^I}\frac{\p V}{\p\bar{w}^{\bar J}} \right\},
\ee
or, using the real fields,
\be \label{eq:epsilon}
\e \geq \frac{1}{4 V^2}\left\{ 4 x^2 \lp \lp \frac{\p V}{\p x} \rp^2 + \lp \frac{\p V}{\p y} \rp^2 \rp + \tilde{K}^{ij} \lp \frac{\p V}{\p
k^i}\frac{\p V}{\p k^j} \rp + q^2 \lp \frac{\p V}{\p q} \rp^2 + s^2 \lp \frac{\p V}{\p s} \rp^2 \right\},
\ee
where we only spelled out the contributions relevant for us and have neglected a positive semi-definite contribution from the axions and the
$\mathcal{U}^A$ and $\mathcal{V}^I$. \\
For several no-go theorems we will use the following inequality (see appendix \ref{app:Kaehler})
\be
\tilde{K}^{ij} \frac{\p V}{\p k^i}\frac{\p V}{\p k^j} = \lp 2 k^i k^j \rp \frac{\p V}{\p k^i}\frac{\p V}{\p k^j} + \lp 2k^i k^j  - 2
(\tilde{X}^{-1})^{ij} \tilde{X}_{kl} k^k k^l \rp \frac{\p V}{\p k^i}\frac{\p V}{\p k^j} \geq 2 \lp k^i \frac{\p V}{\p k^i} \rp^2.
\ee
With these explicit expressions for $V$ and $\e$ we can now derive several new no-go theorems in the spirit of \cite{Flauger:2008ad,Caviezel:2008tf}. All the no-go theorems that involve the $\tau$ modulus will come in pairs, because we can always find a related no-go theorem that uses the new coordinate $\tau' = x'+ i y' \equiv \frac{1}{\tau}$. Since the scalar potential depends only through $M_{ab}$ on $\tau$, its form in terms of $M_{ab}$ is unchanged, and the matrix $M_{ab}$ has to be written in terms of $\tau'$ as
\be
(M_{ab})= \frac{1}{\text{Re}(\tau')} \left(
    \begin{array}{cc}
      1 & -\text{Im}(\tau') \\
      -\text{Im}(\tau') & |\tau'|^2 \\
    \end{array}
  \right),
\ee
so that now
\be
(M)_{ab} A^a A^b = \frac{1}{x'} \ls(x')^2 (A^2)^2 + (A^1 - A^2 (y'))^2 \rs\geq 0,
\ee
since metric positivity required $x>0$ which then implies that $x' =\frac{x}{x^2+y^2} >0$. The form of $\e$ does not change under this coordinate
transformation, since
\be
4 x^2 \lp \lp \frac{\p V}{\p x} \rp^2 + \lp \frac{\p V}{\p y} \rp^2 \rp = 4 (x')^2 \lp \lp \frac{\p V}{\p x'} \rp^2 + \lp \frac{\p V}{\p y'} \rp^2
\rp.
\ee
$M_{ab}$ written in terms of $\tau'$ can be brought into the form \eqref{eq:tau} if we exchange $a,b=1$ and $a,b=2$ everywhere and  simultaneously change the sign of $y'$
\be
(M_{ab}) \stackrel{\substack{1 \leftrightarrow 2\\y' \rightarrow -y'}}{\longrightarrow} (M_{ab}')= \frac{1}{\text{Re}(\tau')} \left(
    \begin{array}{cc}
      |\tau'|^2 & \text{Im}(\tau') \\
      \text{Im}(\tau') & 1 \\
    \end{array}
  \right). \nn\\
\ee
Therefore, for every no-go theorem involving $x$ and $y$, we can find a corresponding one involving $x'$ and $-y'$, if we exchange $a,b=1$ and $a,b=2$.

\subsubsection{The IIB version of the HKTT no-go theorem}
We start out by re-deriving, for our type IIB compactifications, a no-go theorem that has appeared in the context of type IIA flux compactifications on $CY_3$ manifolds in \cite{Hertzberg:2007wc}. There the authors showed that, using the overall volume and dilaton modulus, there is a lower bound $\e \geq\frac{27}{13}$ if one allows for RR-fluxes, $H_3$-flux and O6-planes. Since our setup is formally T-dual to a type IIA compactification with
O6-planes one can ask how this no-go theorem can be translated to our setting. Since $H_3$-flux and metric fluxes get mixed under the formal
T-duality, one expects that we need to restrict the $H_3$-flux and metric flux in our setting. Indeed, if we demand that
\begin{equation}
h^{1A} = r_2^I =
\tilde{r}^{aA}_i = 0 \label{neuegleichung}
\end{equation}
(recall that $\hat{r}_A^{ai} = -\hat{X}_{AB} ( \tilde{X}^{-1} )^{ij} \tilde{r}_j^{aB}$) then we find that
\be
(-3 q \partial_q -x \partial_x - k^i \partial_{k^i}) V \geq 9V,
\ee
where $-x \partial_x - k^i \partial_{k^i} = -\rho \partial_\rho$ with $\rho = (vol_6)^{1/3} = \lp -x\hlf \tilde{X}_{ij} k^i k^j \rp^{1/3}$. From the
explicit expression for $\e$ \eqref{eq:epsilon} we have
\ba
\e &\geq&  \frac{1}{4 V^2}\left\{ 4 x^2 \lp \frac{\p V}{\p x} \rp^2 + 2 \lp k^i \frac{\p V}{\p k^i} \rp^2 + q^2 \lp \frac{\p V}{\p q} \rp^2 \right\}
\nn \\
&=& \frac{1}{V^2} \left\{ \frac{1}{39} \ls (-3 q \partial_q -x \partial_x - k^i \partial_{k^i}) V \rs^2 + \right.\nn \\
&& \left. + \frac{1}{52} \ls (q \partial_q -4(x \partial_x +k^i \partial_{k^i}) V\rs^2 + \frac{1}{6} \ls (2 x \partial_x - k^i \partial_{k^i}) V\rs^2 \right\}\nn\\
&\geq& \frac{1}{V^2} \left\{ \frac{1}{39} \ls (-3 q \partial_q -x \partial_x - k^i \partial_{k^i}) V \rs^2 \right\} \geq \frac{27}{13}. \nn
\ea
Similarly, we can introduce the new variable $\tau' \equiv \frac{1}{\tau}$ as discussed above and demand that $h^{2A} = r_1^I = \tilde{r}^{aA}_i =0$
from which we find
\be
(-3 q \partial_q -x' \partial_{x'} - k^i \partial_{k^i}) V \geq 9V,
\ee
and therefore again $\e \geq \frac{27}{13}$.

We have chosen the assumption (\ref{neuegleichung}) such that a formal T-duality along the 1-direction (c.f. \eqref{eq:T1})  would lead to a manifold without  geometric and non-geometric fluxes in IIA. As was discussed in \cite{Silverstein:2007ac,Caviezel:2008ik,Haque:2008jz}, this no-go theorem can be extended on the type IIA side by allowing certain metric fluxes, i.e. non-Ricci flat manifolds: as long as the resulting compact space has positive curvature everywhere in moduli space, the no-go theorem of \cite{Hertzberg:2007wc} is still applicable. We can likewise ask which of our restrictions (\ref{neuegleichung}) on the fluxes are really needed in the IIB version. From
\ba
&&(-3 q \partial_q -x \partial_x - k^i \partial_{k^i}) V_H \\
&=& 9V_H - 2 \frac{s^2}{16 (-\tilde{X}_{ij} k^i k^j) \, q^2} \hat{K}^{AB} \hat{X}_{AC} \hat{X}_{BD} \lp h^{1C} - \tilde{r}^{1C}_i b^i \rp \lp h^{1D} -
\tilde{r}^{1D}_i b^i \rp,\nn
\ea
we see that we have to require that $h^{1C} - \tilde{r}^{1C}_i b^i=0$ everywhere in moduli space, i.e. for all values of $b^i$. This leads to $h^{1A}=
\tilde{r}^{1A}_i=0$. Using this and making the extra assumption $\tilde{r}^{2A}_i=0$ we find
\be
(-3 q \partial_q -x \partial_x - k^i \partial_{k^i}) V_f = 9 V_f + 2 \frac{1}{16 (-\tilde{X}_{ij} k^i k^j) \, q^2} \frac{1}{s^2} r_2^I r_2^J
\bar{X}_{IK} \bar{X}_{JL} \bar{K}^{KL} \geq 9 V_f.
\ee
So we see that $V_f$ satisfies the no-go theorem even when $r_2^I \neq 0$. It is easy to see that the same is true for all other contributions to the
scalar potential so we conclude that the no-go theorem applies to all models that satisfy only $h^{1A}= \tilde{r}^{aA}_i=0$. In a concrete model it is
also possible to relax the condition $\tilde{r}^{2A}_i=0$ if the model still satisfies
\be
(-3 q \partial_q -x \partial_x - k^i \partial_{k^i}) V_f \geq 9 V_f.
\ee
Again a similar conclusion can be reached considering $x'$ and exchanging $a=1$ and $a=2$ in the previous discussion.

\subsubsection{Factorization in the K\"ahler sector}
Based on our factorization of the K\"ahler moduli into $\tau$ and the $k^i$ we have two more obvious directions in moduli space along which we could
look for no-go theorems \cite{Flauger:2008ad}. If one can show that
\be
-2 q \partial_q V - k^i \partial_{k^i} V \geq 6 V,
\ee
which is certainly true if $\tilde{r}^{aA}_i $ (and hence also $\hat{r}^{aj}_A$ via \eqref{hattilde}) vanish, then one finds from the explicit expression for $\e$ that
\ba
\e &\geq&  \frac{1}{4 V^2}\left\{2 \lp k^i \frac{\p V}{\p k^i} \rp^2 + q^2 \lp \frac{\p V}{\p q} \rp^2 \right\} \nn \\
&=& \frac{1}{V^2} \left\{ \frac{1}{18} \ls (-2 q \partial_q - k^i \partial_{k^i}) V \rs^2 + \frac{1}{36} \ls (-q \partial_q + 4 k^i \partial_{k^i})
V\rs^2 \right\}\\
&\geq& \frac{1}{V^2} \left\{ \frac{1}{18} \ls (-2 q \partial_q - k^i \partial_{k^i}) V \rs^2 \right\} \geq 2. \nn
\ea
As was pointed out in \cite{Caviezel:2008tf} the assumption $\tilde{r}^{aA}_i =0$ can be relaxed, and it is sufficient to show that any particular
model satisfies
\be\label{eq:ep2nogo}
-2 q \partial_q V_f - k^i \partial_{k^i} V_f \geq 6 V_f,
\ee
in order to exclude the existence of dS vacua and the possibility of slow-roll inflation.

Another possible no-go theorem arises if
\be\label{eq:ep95nogo}
-q \partial_q V - x \partial_{x} V \geq 3 V,
\ee
which is always satisfied if $h^{1A} = \tilde{r}_i^{1A} = r_2^I =0$. In this case it does not seem possible to relax these constraints. We find a
lower bound on the slow-roll parameter
\ba
\e &\geq&  \frac{1}{4 V^2}\left\{4 x^2 \lp \frac{\p V}{\p x} \rp^2 + q^2 \lp \frac{\p V}{\p q} \rp^2 \right\} \nn \\
&=& \frac{1}{V^2} \left\{ \frac{1}{5} \ls (-q \partial_q - x \partial_{x}) V \rs^2 + \frac{1}{20} \ls (-q \partial_q + 4 x \partial_{x}) V\rs^2
\right\}\\
&\geq& \frac{1}{V^2} \left\{ \frac{1}{5} \ls (-q \partial_q - x \partial_{x}) V \rs^2 \right\} \geq \frac95. \nn
\ea
Since this no-go theorem involves the $x$-direction we again have a related no-go theorem involving $x'$, if $h^{2A} = \tilde{r}_i^{2A} = r_1^I =0$.

\subsubsection{Factorization in the K\"ahler and complex structure sector}
Just as in \cite{Flauger:2008ad} we can make use of the factorization in the K\"ahler sector and at the same time of the factorization in the complex
structure sector, where we have the two sets of moduli $z^A$ and $w^I$. This can easily be done by looking at directions that involve both $q$ and $s$.
For example, if $h^{1A}=\tilde{r}_i^{1A} = r_a^I = 0$, one finds that
\be
(- 2 q \p_q + s \p_s - x \p_x) V \geq 7 V.
\ee
Using this we can derive a bound on $\e$
\ba
\e &\geq&  \frac{1}{4 V^2}\left\{4 x^2 \lp \frac{\p V}{\p x} \rp^2 + q^2 \lp \frac{\p V}{\p q} \rp^2 + s^2 \lp \frac{\p V}{\p s} \rp^2 \right\} \nn
\\
&=& \frac{1}{V^2} \left\{ \frac{1}{21} \ls (-2 q \p_q + s \p_s - x \p_x) V \rs^2 +\frac{1}{420} \ls (-2q \p_q + s \p_s + 20 x \p_x) V \rs^2 \right.
\\
&& \left.+ \frac{1}{20} \ls (q \partial_q +2 s\partial_{s}) V\rs^2 \right\} \geq \frac{7}{3}. \nn
\ea
Again there is a related no-go theorem involving $x'$ under the assumption that $h^{2A}=\tilde{r}_i^{2A} = r_a^I = 0$.\\
We can also find a no-go theorem using only the complex structure moduli. If $r_a^I=0$, then we have
\be
(- q \p_q + s \p_s) V \geq 4 V,
\ee
and find the following bound on $\e$
\ba
\e &\geq&  \frac{1}{4 V^2}\left\{q^2 \lp \frac{\p V}{\p q} \rp^2 + s^2 \lp \frac{\p V}{\p s} \rp^2 \right\} \nn
\\
&=& \frac{1}{V^2} \left\{ \frac{1}{8} \ls (- q \p_q + s \p_s) V \rs^2 + \frac{1}{8} \ls (q \p_q + s \p_s ) V \rs^2 \right\} \geq 2.
\ea

\subsubsection{No-go theorems for a complete factorization in the K\"ahler and complex structure sector}\label{completefactorization}
In our concrete models the volume does not only factor into $x$ and $\tilde{X}_{ij} k^i k^j$, but we actually have $vol_6 = x k^1 k^2$. This means that we can also study directions involving $k^1$ or $k^2$ or a combination of $x$ with $k^1$ or $k^2$. This leads to no-go theorems very similar to the ones discussed above. Furthermore, we can use the fact that $\hat{X}_{AB} \mathcal{U}^A \mathcal{U}^B \bar{X}_{IJ} \mathcal{V}^I \mathcal{V}^J = -4 \mathcal{U}^1 \mathcal{U}^2 \mathcal{V}^1 \mathcal{V}^2$ to study other directions that generalize the above discussion where we used $q$ and $s$. For each of the cases the restriction we have to put on the NSNS fluxes vary, but we still find the same bound on $\e$. There are of course many more no-go theorems one can derive for these concrete models. We will just discuss two more in detail since we will need them to analyze our explicit models. Since $\hat{X}_{12} =\hat{X}_{21} =-1$ and $\bar{X}_{12} = \bar{X}_{21} =1$ we have
\ba
\e &\geq& V^{-2}\left\{K^{A\bar B}\frac{\p V}{\p z^A} \frac{\p V}{\p\bar{z}^{\bar B}} + K^{I\bar J}\frac{\p V}{\p w^I}\frac{\p V}{\p\bar{w}^{\bar J}}
\right\} \\
&\geq& V^{-2}\left\{ \lp u^A \rp^2 \lp \frac{\p V}{\p u^A} \rp^2  +\lp v^I \rp^2 \lp \frac{\p V}{\p v^I} \rp^2 \right\}.\nn
\ea
It will be convenient to use the $u^A$ and $v^I$. Their explicit dependence can be read off from
\eqref{eq:scalarpotential2}-\eqref{eq:scalarpotential3} if we change $\mathcal{U}^A \rightarrow \frac{u^A}{q \,s}$, $\mathcal{V}^I \rightarrow
\frac{v^I q}{s}$ and $q^4 \rightarrow -u^1 u^2 v^1 v^2$. After this replacement we see that for example
\be
-u^1 \p_{u^1} V \geq V,
\ee
if $\tilde{r}_i^{a2}=h^{a2}=0$. Therefore, we find the bound
\be
\e \geq V^{-2}\left\{ \lp u^1 \rp^2 \lp \frac{\p V}{\p u^1} \rp^2 \right\} \geq 1.
\ee
A similar statement applies for $v^1$ under the assumptions $r_a^2=0$. In both cases there are related no-go theorems if we exchange $A=1$ and $A=2$,
or $I=1$ and $I=2$, respectively.

\subsection{Analysis of concrete examples}
We can now analyze our concrete examples and check whether the existence of dS vacua and slow-roll inflation is not possible due to one of the no-go
theorems from the previous subsection.\\\\
$\mathbf{\frac{SU(3) \times U(1)}{SU(2)} :}$\\
 For this model we have $f^1_{35}=-f^1_{46}=\sqrt{3}/2$, cyclic. Calculating $V_f$ we find
\be
V_f = \frac{3 \mathcal{V}(\mathcal{V}+8 x\, k \, s^2 \mathcal{U})}{16 x\, k^2s^2q^2}.
\ee
Recalling that we need $k>0$ and $\mathcal{U V}<0$ so that the metric is positive definite, we find that
\be
-2q\p_q V_f - k \p_k V_f =6 V_f -\frac{3 \mathcal{U V}}{2 k q^2} \geq 6 V_f
\ee
and therefore that the condition in \eqref{eq:ep2nogo} is satisfied and one has $\e \geq 2$. Under a formal T-duality along the cycle dual to
$Y_2^{(1-+)}$ one obtains a geometric type IIA compactification on the same manifold, which was studied in \cite{Caviezel:2008tf}, where the authors
found the same bound on $\e$.\\\\
$\mathbf{\frac{SU(2)^2}{U(1)} \times U(1) :}$\\
For this model we have $f^1_{35}=1$, cyclic, and $f^1_{46}=-1$. This leads to
\be
V_f = \frac{(\mathcal{V}^1)^2+4 x \, k s^2 \mathcal{U}(\mathcal{V}^1-\mathcal{V}^2)+(\mathcal{V}^2)^2}{8 x\, k^2s^2q^2}.
\ee
Recalling that we need $k>0$, $\mathcal{U V}^1<0$ and $\mathcal{U V}^2>0$ so that the metric is positive definite, we find that
\be
-2q\p_q V_f - k \p_k V_f =6 V_f +\frac{-\mathcal{U} \mathcal{V}^1+\mathcal{U} \mathcal{V}^2}{2 k q^2} \geq 6 V_f
\ee
and therefore that the condition in \eqref{eq:ep2nogo} is satisfied and one again has $\e \geq 2$. Under a formal T-duality along the cycle dual to
$Y_2^{(1-+)}$ one obtains a geometric type IIA compactification on the same manifold, which was studied in \cite{Caviezel:2008tf}, where the authors
found the same bound on $\e$.\\\\
$\mathbf{SU(2) \times SU(2) :}$\\
In \cite{Caviezel:2008tf,Flauger:2008ad} it was shown that compactifications of type IIA on this space can lead to dS extrema with one tachyonic direction. In \cite{deCarlos:2009qm} (based on the earlier work \cite{Font:2008vd,Guarino:2008ik,deCarlos:2009fq}) the authors found fully stabilized dS vacua for type IIA on an SU(3)-structure space with non-geometric fluxes that is very similar to the formally T-dual version of the
SU(2)$\times$SU(2) model studied in the present paper. It is therefore very interesting to ask whether this space allows for geometric dS minima in IIB. We have analyzed the corresponding scalar potential using Mathematica and with the aid of the package STRINGVACUA \cite{Gray:2008zs} but due to its complexity we were only able to find one particular solution with numerically vanishing $\e$. For the ease of presentation we have rounded the values of our solution to six significant digits\footnote{The solution we found has $\e \approx 10^{-20}$. The rounded values in equation \eqref{eq:values} give only $\epsilon \approx 6.5 \times 10^{-3}$.}
\ba \label{eq:values}
&& x \approx0.267585, \quad k^1 \approx 1.76189, \quad k^2 \approx 1.97367,\nn\\
&& u^1 \approx2.38469, \quad u^2 \approx0.0406036, \quad v^1 \approx -0.00820371, \quad v^2 \approx 0.0512969, \nn\\
&& y \approx 0.624470, \quad b^1 \approx -6.22664, \quad b^2 \approx -3.41528,\\
&& c_{(2)}^1 \approx 4.99938, \quad c_{(2)}^2 \approx 6.53845, \quad c_{(4)}^1\approx18.2884, \quad c_{(4)}^2 \approx-14.8650,\nn\\
&& m^1 \approx1.26529, \quad m^2 \approx -1.92725, \quad f^{(3)1} \approx-6.09473, \quad f^{(3)2} \approx 10.5444. \nn
\ea
This solution has $\eta \approx -3.1$ similar to the numerical type IIA dS extrema found in \cite{Caviezel:2008tf,Flauger:2008ad}. Besides this tachyonic direction there is another tachyonic direction corresponding to an eigenvalue of the $\eta$ matrix of approximately -0.00039. While the above solution is not in a regime in which we can trust supergravity, it nevertheless shows that for this model there cannot exist a no-go theorem similar to the ones discussed earlier in this section.

It would be very interesting to study the $\text{SU(2)}\times\text{SU(2)}$ model further to check whether one can prove that there is always at least one tachyonic direction or whether it allows for metastable dS vacua with large volume and small string coupling. Understanding the tachyonic directions better should also allow to decide whether there are points in the moduli space that allow for slow-roll inflation in this model.\\\\
$\mathbf{SU(2) \times U(1)^3 :}$\\
For this model we have $f^1_{35}=1$, cyclic. This leads in terms of $\tau' = \frac{1}{\tau}$ to
\ba
V_f &=& \frac{1}{8 q^2 s^2 x' k^1 k^2} \left\{(k^1 s^2 \mathcal{U}^1)^2 + (k^2 s^2 \mathcal{U}^2)^2 + \left((x')^2+(y')^2\right) (\mathcal{V}^2)^2
\right.\\
&& \left.- 2 (k^1 s^2 \mathcal{U}^1) (k^2 s^2 \mathcal{U}^2) - 2 x' \mathcal{V}^2(k^1 s^2 \mathcal{U}^1+k^2 s^2 \mathcal{U}^2) \right\}.\nn
\ea
We recall that we need $x'>0$, $k^i>0$, $\mathcal{U}^1 \mathcal{U}^2>0$ and $\mathcal{U}^A \mathcal{V}^2>0$ so that the metric is positive definite.
Then we can assume that one of the three quantities $|x' \mathcal{V}^2|$, $|k^1 s^2 \mathcal{U}^1|$, $|k^2 s^2 \mathcal{U}^2|$ is the biggest. We will
choose $|x' \mathcal{V}^2| \geq |k^1 s^2 \mathcal{U}^1|$ and $|x' \mathcal{V}^2| \geq |k^2 s^2 \mathcal{U}^2|$ which leads to
\be
-2q\p_q V_f - k^i \p_{k^i} V_f =6 V_f + \frac{- (k^1 s^2 \mathcal{U}^1-k^2 s^2 \mathcal{U}^2)^2 + x'\mathcal{V}^2 (k^1 s^2 \mathcal{U}^1+k^2 s^2
\mathcal{U}^2)}{4 q^2 s^2 x' k^1 k^2} \geq 6 V_f
\ee
and therefore the condition in \eqref{eq:ep2nogo} is satisfied and one has $\e \geq 2$. One can reach a similar conclusion, if for example $|k^{1} s^2
\mathcal{U}^{1}|$ is the biggest quantity, by looking at the direction $-2q\p_q - k^{2} \p_{k^{2}} -x' \p_{x'}$. Under a formal T-duality along the
cycle dual to $Y_2^{(1-+)}$ one obtains a geometric type IIA compactification on the same manifold, which was studied in \cite{Caviezel:2008tf}, where
the authors found the same bound on $\e$.\\\\
\textbf{The twisted} $\mathbf{T^2 \times T^4/\mathbb{Z}_2 :}$\\
The particular twisted tori models discussed above all satisfy the conditions for one no-go theorem similar to the ones explicitly discussed in section \ref{nogo} (see in particular the discussion in subsection \ref{completefactorization}). From the possible $H_3$- and metric fluxes given in \eqref{eq:NSNSfluxes} one finds
\begin{center}
\begin{tabular}{|c|c|}
  \hline
  no-go theorem & model \\ \hline\hline
  $\e \geq \frac73$ & $\begin{array}{c}
                        \{1\}, \{3\}, \{3,5\}, \{1,2,3,4\}, \{1,2,3,5\}, \{1,2,3,6\}, \\
                        \{3,4,5,6\}, \{1,2,3,4,5\}, \{1,3,4,5,6\}
                      \end{array}$ \\\hline
  $\e \geq 2$ & $\{1,2\}$, $\{1,2,3\}$, $\{1,3,4\}$, $\{1,3,5\}$, $\{1,3,6\}$, $\{3,4,5\}$,  $\{1,3,4,5\}$ \\\hline
  $\e \geq \frac95$ & $\{3,6\}$ \\\hline
  $\e \geq 1$ & $\{1,3\}$, $\{3,4\}$ \\
  \hline
\end{tabular}
\end{center}
where again $\{\ldots\}$ denote the chosen base over which we twisted the complementary part of the torus. Unlike the situation for the coset spaces, where only the SU(2)$\times$SU(2)-space T-dualized to a non-geometric compactification in type IIA, the twisted tori in the above table generically become non-geometric spaces in IIA, and the no-go theorems proven here are hence in general not covered by the no-go theorems proven in \cite{Flauger:2008ad}. The no-go theorem corresponding to $\epsilon \geq 1$ also makes use of a different direction in moduli space as the no-go theorems considered in \cite{Flauger:2008ad}.

It is also possible to allow combinations of NSNS-fluxes that cannot be explicitly constructed from the base-fiber splitting. Then it is not a priori
clear that there is an actual compact space that corresponds to such metric fluxes. However, there are certainly examples where this is the
case like for example SU(2)$\times$SU(2), which cannot be constructed from a base-fiber splitting of a torus but which is nevertheless a good
geometric compactification spaces. In such a case one can hope to evade all no-go theorems related to the $\e$ parameter and find dS extrema in the
potential. This was shown to be the case in \cite{Flauger:2008ad} in type IIA compactifications on SU(3)-structure manifolds, where the authors found
(unstable) dS extrema by using metric fluxes that can be obtained from a base-fiber splitting but also had examples of dS extrema that could not be constructed in this way.

\section{Conclusion}\label{conclusion}
In this paper we have studied type IIB compactifications on six-dimensional SU(2)-structure manifolds in the presence of O5- and O7-planes, $H_3$- and RR-fluxes. We have spelled out the resulting classical four-dimensional $\mathcal{N}=1$ supergravity action and studied the scalar potential for the closed string moduli. We have shown that it is possible to stabilize all bulk closed string moduli at tree-level in supersymmetric AdS vacua with large volume and small string coupling, which we made fully explicit for a specific model. We found that, by contrast, supersymmetric Minkowski vacua are not possible in this setup.  We discussed many  explicit examples of six-dimensional SU(2)-structure manifolds in detail and discussed their cosmological properties. We derived potential no-go theorems against dS vacua and slow-roll inflation for our class of models and were able to use them to exclude all but one of our explicit models. However, there is no known generic no-go theorem that forbids dS vacua or slow-roll inflation in this class of models, which makes it interesting to study further models. Also for one of our explicit models we found a dS solution with numerically vanishing $\epsilon$ and two tachyonic directions. It would be interesting to study this further in particular since it was shown in \cite{deCarlos:2009qm} that it is possible to obtain fully stabilized dS vacua in non-geometric compactifications of type IIA on SU(3)-structure spaces, and our setup is formally T-dual to compactifications of this type.\\
A straightforward extension of our work would be to consider type IIB on SU(2)-structure manifolds and include only one kind of orientifold planes. In that case it would be interesting to find out whether it is still possible to stabilize all closed string moduli at tree-level in the resulting four-dimensional $\mathcal{N}=2$ theory (see \cite{GomezReino:2008bi,Roest:2009dq,Dall'Agata:2009gv} for related work on dS vacua in flux compactifications with $\mathcal{N}>1$ SUSY). It would also be interesting to extend the work of \cite{Danckaert:2009hr} and study type IIA compactifications on SU(2)-structure manifolds in the presence of two different kinds of O-planes. One can also analyze our setup from a 10-dimensional point of view as was done for type IIA on SU(3)-structure manifolds in \cite{Danielsson:2009ff}. We hope to come back to some of these issues in the future.

\begin{acknowledgments}
We would like to thank Martin Ammon, Ralph Blumenhagen, Davide Cassani, Gianguido Dall'Agata, Beatriz de Carlos, Johanna Erdmenger, Raphael Flauger, Thomas Grimm, Christian Gross, Adolfo Guarino, Paul Koerber, Simon K\"ors, Jan Louis, Dieter L\"ust, Danny Mart\'{i}nez-Pedrera, Jesus Moreno, Eran Palti, Maximilian Schmidt-Sommerfeld and Thomas Van Riet for useful discussions. We are especially grateful to Daniel Robbins for reading and commenting on the manuscript. This work was supported by the German Research Foundation (DFG) within the Emmy Noether Program (Grant number ZA 279/1-2) and the Cluster of Excellence "Center for Quantum Engineering and Spacetime Research (QUEST)".
\end{acknowledgments}

\begin{appendix}

\section{D-terms in type IIB} \label{app:DtermsIIB}
In this appendix we discuss the additional terms in the scalar potential that arise, if the SU(2)-structure manifold has forms $Y^{(2++)}_\a$ that are even under both the O5- and the O7-orientifold projections.\\
We define the additional symmetric intersection form
\be
\breve{X}_{\a\b} = \int Y^{(1-+)}_1 \wedge Y^{(1-+)}_2 \wedge Y^{(2++)}_\a \wedge Y^{(2++)}_\b.
\ee
Acting with $\dd$ on the forms can now lead to new contributions for $Y^{(2-+)}_I$ (c.f. \eqref{eq:IIBd}) and we also allow for the non-closure of the $Y^{(2++)}_\a$ forms
\ba
\dd Y^{(2-+)}_I &=& \bar{r}^{a\a}_I Y^{(1-+)}_a \wedge Y^{(2++)}_\a, \nn\\
\dd Y^{(2++)}_\a &=& \breve{r}^{aI}_\a Y^{(1-+)}_a \wedge Y^{(2-+)}_I. \nn
\ea
The two new matrices are not independent but rather satisfy $\breve{r}^{aI}_\a = -\breve{X}_{\a\b} \lp \bar{X}^{-1} \rp^{IJ} \bar{r}^{a\b}_J$. Demanding that $\dd$ squares to zero on the forms one finds the necessary conditions (c.f. \eqref{eq:dsquaredIIB})
\be\label{eq:dsquared2}
\bar{X}_{IJ} r^I_a \breve{r}_\a^{bJ} = \e_{ab} \tilde{r}_i^{aA} \hat{r}_A^{bj} = \e_{ab} \hat{r}_A^{ai} \tilde{r}_i^{bB} = \e_{ab} \bar{r}_I^{a\a} \breve {r}_\a^{bJ} = \e_{ab} \breve{r}_\a^{aI} \bar{r}_I^{b\b} =0.
\ee
From the RR form $C_4$ we now also get $U(1)$ vector fields
\be
C_4 = A^{a\a}_\mu \dd x^\mu \w Y^{(1-+)}_a \wedge Y^{(2++)}_\a.
\ee
Note that due to the self-duality of $F_5$ not all of these gauge fields are independent. Half of them are the magnetic duals of the other half.\\
The four-dimensional action is uniquely determined in terms of three functions. The K\"ahler potential $K$ \eqref{eq:Kaehlerpotential}, \eqref{eq:Kaehlerpotential2}, the holomorphic superpotential $W$ \eqref{eq:superpotential} and the holomorphic gauge-kinetic coupling $f_{(a\a)(b\b)}$ \footnote{We will stick to our notation in which the different $U(1)$ gauge fields are distinguished by two different indices $a$ and $\al$. To increase legibility we will group the four indices on the gauge-kinetic function in groups of two indices by using parenthesis.}. The bosonic part of the action is
\ba
S^{(4)} &=& -\int_{M^4}\left\{-\hlf R\ast 1+K_{M\bar N}d\phi^M\w\ast d\bar\phi^{\bar N}+V\ast 1\right.\non\\
&& \left.+\hlf \text{Re}( f_{(a\al)(b\beta)}) F^{a\al}\w\ast F^{b\beta}+\hlf \text{Im} (f_{(a\al)(b\beta)}) F^{a\al}\w F^{b\beta}\right\},
\ea
where $\ast$ is the four-dimensional Hodge star, $K_{M\bar N}=\p_M\bar\p_{\bar N}K$, $K^{M\bar N}$ is its transposed inverse, $F^{a\al}=\dd A^{a\al}$, and $D_M W=\p_M W+(\p_M K)W$ and $M,N$ run over all scalar fields, i.e. over $\{a,i,A,I\}$. Our main interest is in the scalar potential
\be
V= e^K \lp K^{M\bar N}D_MW\,\overline{D_{ N} W}-3|W|^2\rp+\hlf\lp\text{Re}(f)\rp^{-1\,(a\al)(b\beta)}D_{a\a} D_{b\b}.
\ee
The D-terms $D_{a\a}$ for the $U(1)$ gauge groups coming from reducing $C_4$ are
\be\label{eq:Dterm}
D_{a\al}= \frac{i}{W}\d_{a\al}\phi^M D_M W = i\p_M K \d_{a\al} \phi^M + i\frac{\d_{a\al} W}{W},
\ee
where $\d_{a\al}\phi^M$ and $\d_{a\al} W$ are the variations of the field $\phi^M$ and superpotential $W$ under a gauge transformation. The equation \eqref{eq:Dterm} is not valid for $W=0$ but our compactification will have $\d_{a\al} W=0$, so that the explicit D-terms determined below are also valid for $W=0$.\\
The holomorphic gauge-kinetic coupling can be read off directly from the 10-dimensional action. See \cite{Grimm:2004uq} for an explicit derivation for type IIB on SU(3)-structure manifolds.\\
We have seen that the RR field $C_4$ gives rise to $U(1)$ gauge fields and axions through the expansion
\be
C_4 = A^{a\a}_\mu \dd x^\mu \w Y^{(1-+)}_a \wedge Y^{(2++)}_\a +c_{(4)}^I Y^{(1-+)}_1 \wedge Y^{(1-+)}_2 \wedge Y^{(2-+)}_I.
\ee
Doing a gauge transformation of $C_4$ we find
\ba
&&C_4 \rightarrow C_4 + \dd \Lambda^{(3-+)} \nn\\
&=& C_4 + \dd \lp \lambda^{a\a}(x) \, Y^{(1-+)}_a \wedge Y^{(2++)}_\a\rp \nn \\
&=& C_4 +  \p_\mu \lambda^{a\a} \, \dd x^\m \w  Y^{(1-+)}_a \wedge Y^{(2++)}_\a - \e_{ab} \lambda^{a\a} \breve{r}_\a^{bI} \, Y^{(1-+)}_1 \wedge Y^{(1-+)}_2 \wedge Y^{(2-+)}_I \\
&=& \lp A^{a\a}_\mu +\p_\mu \lambda^{a\a} \rp \dd x^\mu \w Y^{(1-+)}_a \wedge Y^{(2++)}_\a + \lp c_{(4)}^I - \e_{ab} \lambda^{a\a} \breve{r}_\a^{bI} \rp \,Y^{(1-+)}_1 \wedge Y^{(1-+)}_2 \wedge Y^{(2-+)}_I \nn,
\ea
so we see that $A^{a\a}$ transforms as a $U(1)$ gauge field and that the axions $c_{(4)}^I$ also transform, i.e. they carry the charge $-\e_{ab} \breve{r}_\a^{bI}$ under the $A^{a\a}$ gauge group. All other fields are invariant so that we find the following D-terms
\be
D_{a\al}=-i \frac{\p K}{\p w^I} \e_{ab} \breve{r}_\a^{bI}= i \frac{\bar{X}_{IJ} v^J}{\bar{X}_{KL}v^Kv^L} \e_{ab} \breve{r}_\a^{bI},
\ee
where we have used that
\be
\d_{a\al} W= -i T^c \bar{X}_{IJ} \lp -i \e_{ab} \breve{r}_\a^{bI} \rp r_c^J = 0,
\ee
due to the first constraint in \eqref{eq:dsquared2}.\\
So the D-term potential is
\be\label{eq:Dtermpot}
V_D = -\hlf\lp\text{Re}(f)\rp^{-1\,(a\al)(b\beta)} \lp \frac{\bar{X}_{IJ} v^J}{\bar{X}_{KL}v^Kv^L} \e_{ac} \breve{r}_\a^{cI} \rp \lp\frac{\bar{X}_{IJ} v^J}{\bar{X}_{KL}v^Kv^L} \e_{bd} \breve{r}_\b^{dI}\rp.
\ee

\section{Base-fiber construction}\label{app:basefiber}
In this appendix we will explain the base-fiber construction \cite{Hull:2004in,Dabholkar:2005ve,Grana:2006kf,Hull:2006qs,Hull:2006va,Cvetic:2007ju,Ihl:2007ah,Bergman:2007qq} using one particular example as illustration. For a more detailed discussion and several explicit examples we refer the reader to \cite{Cvetic:2007ju,Ihl:2007ah}.\\
For the base-fiber construction we split the internal space into a $(6-n)$-dimensional base and an $n$-dimensional fiber. The only requirement for this split is that base and fiber do not mix under the orbifold and orientifold actions. If we know the T-duality group of the fiber and the base space is not simply-connected, then when going around a non-trivial cycle in the base we can twist the fiber by an element of the T-duality group. Since the underlying string theory is invariant under the T-duality group, the resulting space is a valid string compactification. However, in general the resulting space is not globally but only locally geometric. Since we are interested in supergravity compactifications, we will restrict ourselves to geometric twists.\\
We will now study an explicit twist of the space $T^2 \times T^4/\mathbb{Z}_2$ with coordinates $x^p \sim x^p+1, \, p=1,\ldots 6$. The $\mathbb{Z}_2$ acts by
\be
\theta: (x^1,x^2,x^3,x^4,x^5,x^6) \rightarrow (x^1,x^2,-x^3,-x^4,-x^5,-x^6).
\ee
Furthermore, we want to allow for O5- and O7-planes that can be obtained from the projections $\Om_p \s_{O5}$ and $(-1)^{F_L} \Om_p \s_{O7}$ where
\ba
\s_{O5}: &&(x^1,x^2,x^3,x^4,x^5,x^6) \rightarrow (-x^1,-x^2,x^3,x^4,-x^5,-x^6),\\
\s_{O7}: && (x^1,x^2,x^3,x^4,x^5,x^6) \rightarrow (x^1,x^2,x^3,-x^4,x^5,-x^6).
\ea
Now for our concrete example we choose the base to be spanned by $x^b, \, b=\{1,2,3,5\}$ and the fiber by $x^f,\, f=\{4,6\}$. We will use indices $b$ for the base and $f$ for the fiber. Since the fiber is a $T^2$ (moduli the orbifold and orientifold projections which we will take into account below), the T-duality group for the fiber is SO$(2,2;\mathbb{Z})$. Now for each base index we choose an element in the Lie algebra of the T-duality group \footnote{The lower block in the $M_b$ corresponds to non-geometric $Q$-flux which we have set to zero.}
\be
M_b =\left(
       \begin{array}{cc}
         -f^{f_2}_{bf_1} & H_{b f_1 f_2} \\
         0_{2 \times 2} & f^{f_1}_{b f_2} \\
       \end{array}
     \right)
\ee
that corresponds to an infinitesimal twist of the fiber when moving along the circle direction $x^b$. From the action of the T-duality group on the fields one can identify the entries in these matrices with $H_3$-flux and metric flux $f^m_{np}$ \cite{Ihl:2007ah}. To be consistent with the orbifold and orientifold projections, we have to demand that the $f^m_{np}$ are invariant under $\theta, \s_{O5}, \s_{O7}$ while $H_3$ needs to be even under $\theta$ and odd under $\s_{O5}$ and $\s_{O7}$. This gives the following matrices
\be
M_1 = \left(
        \begin{array}{cccc}
          0 & -f^6_{14} & 0 & 0 \\
          -f^4_{16} & 0 & 0 & 0 \\
          0 & 0 & 0 & f^4_{16} \\
          0 & 0 & f^6_{14} & 0 \\
        \end{array}
      \right), \,
M_2 = \left(
        \begin{array}{cccc}
          0 & -f^6_{24} & 0 & 0 \\
          -f^4_{26} & 0 & 0 & 0 \\
          0 & 0 & 0 & f^4_{26} \\
          0 & 0 & f^6_{24} & 0 \\
        \end{array}
      \right), \,
M_3 = M_4 = \left(
        \begin{array}{cccc}
          0 & 0 & 0 & 0 \\
          0 & 0 & 0 & 0 \\
          0 & 0 & 0 & 0 \\
          0 & 0 & 0 & 0 \\
        \end{array}
      \right).
\ee
So we see that this particular setup does not allow $H_3$-flux and we can have at most four different metric fluxes. \\
There are two more requirements the $M_b$ have to satisfy. If we move around a trivial cycle in the base the resulting twist has to be trivial since we can shrink the cycle to zero. This requirement is implemented by demanding that $[M_{b_1},M_{b_2}]=0$ which in our case gives
\be
[M_1, M_2] = \left(
        \begin{array}{cccc}
          -f^4_{16} f^6_{24} + f^4_{26} f^6_{14} & 0 & 0 & 0 \\
          0 & f^4_{16} f^6_{24} - f^4_{26} f^6_{14} & 0 & 0 \\
          0 & 0 & f^4_{16} f^6_{24} - f^4_{26} f^6_{14} & 0 \\
          0 & 0 & 0 & -f^4_{16} f^6_{24} + f^4_{26} f^6_{14} \\
        \end{array}
      \right)=0.
\ee
This constraint is the Bianchi identity \eqref{eq:Bianchi}.\\
The final constraint that arises from the base fiber splitting is that $e^{M_b} \in$ SO$(2,2;\mathbb{Z})$. This constraint gives us the right quantization of the NSNS fluxes in our new space. For the generic case in which $M_b$ is not nil-potent one finds that the NSNS fluxes are not integers but rather real numbers \cite{Cvetic:2007ju,Ihl:2007ah}. It is also possible that the quantization condition forces certain fluxes to vanish, e.g. if one finds that $e^{f^{f_1}_{b f_2}},e^{-f^{f_1}_{b f_2}} \in \mathbb{Z}$. In our example we find
\be
\cosh{\lp \sqrt{f^4_{a6}} \sqrt{f^6_{a4}}\rp}, \, \frac{\sqrt{f^6_{a4}} \sinh{(\sqrt{f^4_{a6}} \sqrt{f^6_{a4}})}}{\sqrt{f^4_{a6}}}, \, \frac{\sqrt{f^4_{a6}} \sinh{(\sqrt{f^4_{a6}} \sqrt{f^6_{a4}})}}{\sqrt{f^6_{a4}}} \in \mathbb{Z}, \, \, a=1,2.
\ee

\section{Useful relations for the calculation of the explicit scalar potential}\label{app:Kaehler}
The type IIB K\"ahler potential (cf. \eqref{eq:Kaehlerpotential}, \eqref{eq:Kaehlerpotential2})
\ba
K &=& - \ln \ls -(\tau + \bar{\tau})\hlf\tilde{X}_{ij} (t^i+\bar{t}^i)(t^j+\bar{t}^j) \rs \\
&& - \ln \ls \frac{1}{64} \hat{X}_{AB} (z^A+\bar{z}^A)(z^B+\bar{z}^B) \bar{X}_{IJ} (w^I+\bar{w}^I)(w^J+\bar{w}^J) \rs \nn
\ea
has the following useful properties
\ba
e^K &=& \frac{1}{8 q^4 vol_6},\nn \\
\p_\tau K &=& -\frac{1}{2x}, \quad \p_{t^i} K = -\frac{\tilde{X}_{ij} k^j}{\tilde{X}_{kl} k^k k^l},\nn \\
\p_{z^A} K &=& -\frac{1}{s\, q} \frac{\hat{X}_{AB}\mathcal{U}^B}{\hat{X}_{CD} \mathcal{U}^C \mathcal{U}^D}, \quad \p_{w^I} K = -\frac{s}{q} \frac{\bar{X}_{IJ} \mathcal{V}^J}{\bar{X}_{KL} \mathcal{V}^K \mathcal{V}^L},\nn \\
K^{\tau \bar{\tau}} &=& 4x^2, \quad K^{i \bar{j}} = 4 k^i k^j - 2 (\tilde{X}^{-1})^{ij} \tilde{X}_{kl} k^k k^l, \nn \\
K^{A \bar{B}} &=& q^2 s^2\lp 4 \mathcal{U}^A \mathcal{U}^B - 2 (\hat{X}^{-1})^{AB} \hat{X}_{CD} \mathcal{U}^C \mathcal{U}^D \rp,\nn \\
K^{I \bar{J}} &=& \frac{q^2}{s^2} \lp 4 \mathcal{V}^I \mathcal{V}^J - 2 (\bar{X}^{-1})^{IJ} \bar{X}_{KL} \mathcal{V}^K \mathcal{V}^L \rp, \nn \\
K^{\tau \bar{\tau}} \p_{\bar{\tau}} K &=&-2x, \quad K^{i \bar{j}} \p_{\bar{t}^j} K = - 2k^i, \nn \\
K^{A \bar{B}} \p_{\bar{z}^B} K &=& - 2 q s \, \mathcal{U}^A, \quad K^{I \bar{J}} \p_{\bar{w}^J} K = - \frac{2 q}{s} \mathcal{V}^I, \nn
\ea
where $q=e^{-D}=e^{-\phi} \sqrt{vol_6}$, $vol_6 = -\hlf x \tilde{X}_{ij} k^i k^j$ and Re$(z^A)=q s\,\mathcal{U}^A$, Re$(w^I)=\frac{q}{s} \mathcal{V}^I$ such that $\hat{X}_{AB} \mathcal{U}^A \mathcal{U}^B \bar{X}_{IJ} \mathcal{V}^I \mathcal{V}^J =4$.\\
This K\"ahler potential satisfies the scaling condition
\be
K^{\tau \bar{\tau}} \p_{\tau} K \p_{\bar{\tau}} K + K^{i \bar{j}} \p_{t^i} K \p_{\bar{t}^j} K + K^{A \bar{B}} \p_{z^A} K \p_{\bar{z}^B} K + K^{I \bar{J}} \p_{w^I} K \p_{\bar{w}^J} K = 1 +2 +2 +2 =7.
\ee

\end{appendix}

\bibliographystyle{JHEP}
\bibliography{SU(2)structure}

\providecommand{\href}[2]{#2}\begingroup\raggedright\begin{thebibliography}{10}

\bibitem{Grana:2005jc}
M.~Grana, {\it {Flux compactifications in string theory: A comprehensive
  review}},  {\em Phys. Rept.} {\bf 423} (2006) 91--158
  [\href{http://arXiv.org/abs/hep-th/0509003}{{\tt hep-th/0509003}}].

\bibitem{Douglas:2006es}
M.~R. Douglas and S.~Kachru, {\it {Flux compactification}},  {\em Rev. Mod.
  Phys.} {\bf 79} (2007) 733--796
  [\href{http://arXiv.org/abs/hep-th/0610102}{{\tt hep-th/0610102}}].

\bibitem{Blumenhagen:2006ci}
R.~Blumenhagen, B.~K{\"o}rs, D.~L{{\"u}}st and S.~Stieberger, {\it
  {Four-dimensional String Compactifications with D-Branes, Orientifolds and
  Fluxes}},  {\em Phys. Rept.} {\bf 445} (2007) 1--193
  [\href{http://arXiv.org/abs/hep-th/0610327}{{\tt hep-th/0610327}}].

\bibitem{Denef:2007pq}
F.~Denef, M.~R. Douglas and S.~Kachru, {\it {Physics of string flux
  compactifications}},  {\em Ann. Rev. Nucl. Part. Sci.} {\bf 57} (2007)
  119--144 [\href{http://arXiv.org/abs/hep-th/0701050}{{\tt hep-th/0701050}}].

\bibitem{Gukov:1999ya}
S.~Gukov, C.~Vafa and E.~Witten, {\it {CFT's from Calabi-Yau four-folds}},
  {\em Nucl. Phys.} {\bf B584} (2000) 69--108
  [\href{http://arXiv.org/abs/hep-th/9906070}{{\tt hep-th/9906070}}].

\bibitem{Giddings:2001yu}
S.~B. Giddings, S.~Kachru and J.~Polchinski, {\it {Hierarchies from fluxes in
  string compactifications}},  {\em Phys. Rev.} {\bf D66} (2002) 106006
  [\href{http://arXiv.org/abs/hep-th/0105097}{{\tt hep-th/0105097}}].

\bibitem{Kachru:2003aw}
S.~Kachru, R.~Kallosh, A.~Linde and S.~P. Trivedi, {\it {De Sitter vacua in
  string theory}},  {\em Phys. Rev.} {\bf D68} (2003) 046005
  [\href{http://arXiv.org/abs/hep-th/0301240}{{\tt hep-th/0301240}}].

\bibitem{Becker:2002nn}
K.~Becker, M.~Becker, M.~Haack and J.~Louis, {\it {Supersymmetry breaking and
  alpha'-corrections to flux induced potentials}},  {\em JHEP} {\bf 06} (2002)
  060 [\href{http://arXiv.org/abs/hep-th/0204254}{{\tt hep-th/0204254}}].

\bibitem{Balasubramanian:2004uy}
V.~Balasubramanian and P.~Berglund, {\it {Stringy corrections to Kahler
  potentials, SUSY breaking, and the cosmological constant problem}},  {\em
  JHEP} {\bf 11} (2004) 085 [\href{http://arXiv.org/abs/hep-th/0408054}{{\tt
  hep-th/0408054}}].

\bibitem{Balasubramanian:2005zx}
V.~Balasubramanian, P.~Berglund, J.~P. Conlon and F.~Quevedo, {\it {Systematics
  of Moduli Stabilisation in Calabi-Yau Flux Compactifications}},  {\em JHEP}
  {\bf 03} (2005) 007 [\href{http://arXiv.org/abs/hep-th/0502058}{{\tt
  hep-th/0502058}}].

\bibitem{Louis:2002ny}
J.~Louis and A.~Micu, {\it {Type II theories compactified on Calabi-Yau
  threefolds in the presence of background fluxes}},  {\em Nucl. Phys.} {\bf
  B635} (2002) 395--431 [\href{http://arXiv.org/abs/hep-th/0202168}{{\tt
  hep-th/0202168}}].

\bibitem{Grimm:2004ua}
T.~W. Grimm and J.~Louis, {\it {The effective action of type IIA Calabi-Yau
  orientifolds}},  {\em Nucl. Phys.} {\bf B718} (2005) 153--202
  [\href{http://arXiv.org/abs/hep-th/0412277}{{\tt hep-th/0412277}}].

\bibitem{DeWolfe:2005uu}
O.~DeWolfe, A.~Giryavets, S.~Kachru and W.~Taylor, {\it {Type IIA moduli
  stabilization}},  {\em JHEP} {\bf 07} (2005) 066
  [\href{http://arXiv.org/abs/hep-th/0505160}{{\tt hep-th/0505160}}].

\bibitem{Ihl:2006pp}
M.~Ihl and T.~Wrase, {\it {Towards a realistic type IIA T**6/Z(4) orientifold
  model with background fluxes. I: Moduli stabilization}},  {\em JHEP} {\bf 07}
  (2006) 027 [\href{http://arXiv.org/abs/hep-th/0604087}{{\tt
  hep-th/0604087}}].

\bibitem{Lust:2004ig}
D.~L{\"u}st and D.~Tsimpis, {\it {Supersymmetric AdS(4) compactifications of
  IIA supergravity}},  {\em JHEP} {\bf 02} (2005) 027
  [\href{http://arXiv.org/abs/hep-th/0412250}{{\tt hep-th/0412250}}].

\bibitem{Derendinger:2004jn}
J.-P. Derendinger, C.~Kounnas, P.~M. Petropoulos and F.~Zwirner, {\it
  {Superpotentials in IIA compactifications with general fluxes}},  {\em Nucl.
  Phys.} {\bf B715} (2005) 211--233
  [\href{http://arXiv.org/abs/hep-th/0411276}{{\tt hep-th/0411276}}].

\bibitem{Villadoro:2005cu}
G.~Villadoro and F.~Zwirner, {\it {N = 1 effective potential from dual type-IIA
  D6/O6 orientifolds with general fluxes}},  {\em JHEP} {\bf 06} (2005) 047
  [\href{http://arXiv.org/abs/hep-th/0503169}{{\tt hep-th/0503169}}].

\bibitem{House:2005yc}
T.~House and E.~Palti, {\it {Effective action of (massive) IIA on manifolds
  with SU(3) structure}},  {\em Phys. Rev.} {\bf D72} (2005) 026004
  [\href{http://arXiv.org/abs/hep-th/0505177}{{\tt hep-th/0505177}}].

\bibitem{Grana:2005ny}
M.~Grana, J.~Louis and D.~Waldram, {\it {Hitchin functionals in N = 2
  supergravity}},  {\em JHEP} {\bf 01} (2006) 008
  [\href{http://arXiv.org/abs/hep-th/0505264}{{\tt hep-th/0505264}}].

\bibitem{Camara:2005dc}
P.~G. Camara, A.~Font and L.~E. Ibanez, {\it {Fluxes, moduli fixing and
  MSSM-like vacua in a simple IIA orientifold}},  {\em JHEP} {\bf 09} (2005)
  013 [\href{http://arXiv.org/abs/hep-th/0506066}{{\tt hep-th/0506066}}].

\bibitem{KashaniPoor:2006si}
A.-K. Kashani-Poor and R.~Minasian, {\it {Towards reduction of type II theories
  on SU(3) structure manifolds}},  {\em JHEP} {\bf 03} (2007) 109
  [\href{http://arXiv.org/abs/hep-th/0611106}{{\tt hep-th/0611106}}].

\bibitem{Grana:2006hr}
M.~Grana, J.~Louis and D.~Waldram, {\it {SU(3) x SU(3) compactification and
  mirror duals of magnetic fluxes}},  {\em JHEP} {\bf 04} (2007) 101
  [\href{http://arXiv.org/abs/hep-th/0612237}{{\tt hep-th/0612237}}].

\bibitem{Benmachiche:2006df}
I.~Benmachiche and T.~W. Grimm, {\it {Generalized N = 1 orientifold
  compactifications and the Hitchin functionals}},  {\em Nucl. Phys.} {\bf
  B748} (2006) 200--252 [\href{http://arXiv.org/abs/hep-th/0602241}{{\tt
  hep-th/0602241}}].

\bibitem{Koerber:2007xk}
P.~Koerber and L.~Martucci, {\it {From ten to four and back again: how to
  generalize the geometry}},  {\em JHEP} {\bf 08} (2007) 059
  [\href{http://arXiv.org/abs/0707.1038}{{\tt 0707.1038}}].

\bibitem{Ihl:2007ah}
M.~Ihl, D.~Robbins and T.~Wrase, {\it {Toroidal Orientifolds in IIA with
  General NS-NS Fluxes}},  {\em JHEP} {\bf 08} (2007) 043
  [\href{http://arXiv.org/abs/0705.3410}{{\tt 0705.3410}}].

\bibitem{Cassani:2007pq}
D.~Cassani and A.~Bilal, {\it {Effective actions and N=1 vacuum conditions from
  SU(3) x SU(3) compactifications}},  {\em JHEP} {\bf 09} (2007) 076
  [\href{http://arXiv.org/abs/0707.3125}{{\tt 0707.3125}}].

\bibitem{KashaniPoor:2007tr}
A.-K. Kashani-Poor, {\it {Nearly Kaehler Reduction}},  {\em JHEP} {\bf 11}
  (2007) 026 [\href{http://arXiv.org/abs/0709.4482}{{\tt 0709.4482}}].

\bibitem{Cassani:2008rb}
D.~Cassani, {\it {Reducing democratic type II supergravity on SU(3) x SU(3)
  structures}},  {\em JHEP} {\bf 06} (2008) 027
  [\href{http://arXiv.org/abs/0804.0595}{{\tt 0804.0595}}].

\bibitem{Hertzberg:2007ke}
M.~P. Hertzberg, M.~Tegmark, S.~Kachru, J.~Shelton and O.~Ozcan, {\it
  {Searching for Inflation in Simple String Theory Models: An Astrophysical
  Perspective}},  {\em Phys. Rev.} {\bf D76} (2007) 103521
  [\href{http://arXiv.org/abs/0709.0002}{{\tt 0709.0002}}].

\bibitem{Hertzberg:2007wc}
M.~P. Hertzberg, S.~Kachru, W.~Taylor and M.~Tegmark, {\it {Inflationary
  Constraints on Type IIA String Theory}},  {\em JHEP} {\bf 12} (2007) 095
  [\href{http://arXiv.org/abs/0711.2512}{{\tt 0711.2512}}].

\bibitem{Silverstein:2007ac}
E.~Silverstein, {\it {Simple de Sitter Solutions}},  {\em Phys. Rev.} {\bf D77}
  (2008) 106006 [\href{http://arXiv.org/abs/0712.1196}{{\tt 0712.1196}}].

\bibitem{Haque:2008jz}
S.~S. Haque, G.~Shiu, B.~Underwood and T.~Van~Riet, {\it {Minimal simple de
  Sitter solutions}},  {\em Phys. Rev.} {\bf D79} (2009) 086005
  [\href{http://arXiv.org/abs/0810.5328}{{\tt 0810.5328}}].

\bibitem{Danielsson:2009ff}
U.~H. Danielsson, S.~S. Haque, G.~Shiu and T.~Van~Riet, {\it {Towards Classical
  de Sitter Solutions in String Theory}},  {\em JHEP} {\bf 09} (2009) 114
  [\href{http://arXiv.org/abs/0907.2041}{{\tt 0907.2041}}].

\bibitem{Caviezel:2008tf}
C.~Caviezel, P.~Koerber, S.~K{\"o}rs, D.~L{\"u}st, T.~Wrase and M.~Zagermann,
  {\it {On the Cosmology of Type IIA Compactifications on SU(3)- structure
  Manifolds}},  {\em JHEP} {\bf 04} (2009) 010
  [\href{http://arXiv.org/abs/0812.3551}{{\tt 0812.3551}}].

\bibitem{Flauger:2008ad}
R.~Flauger, S.~Paban, D.~Robbins and T.~Wrase, {\it {Searching for slow-roll
  moduli inflation in massive type IIA supergravity with metric fluxes}},  {\em
  Phys. Rev.} {\bf D79} (2009) 086011
  [\href{http://arXiv.org/abs/0812.3886}{{\tt 0812.3886}}].

\bibitem{Nilsson:1984bj}
B.~E.~W. Nilsson and C.~N. Pope, {\it {Hopf fibration of eleven-dimensional
  supergravity}},  {\em Class. Quant. Grav.} {\bf 1} (1984) 499.

\bibitem{Lust:1986ix}
D.~L{\"u}st, {\it {Compactification Of Ten-Dimensional Superstring Theories
  Over Ricci Flat Coset Spaces}},  {\em Nucl. Phys.} {\bf B276} (1986) 220.

\bibitem{Castellani:1986rg}
L.~Castellani and D.~L{\"u}st, {\it {Superstring compactification on
  homogeneous coset spaces with torsion}},  {\em Nucl. Phys.} {\bf B296} (1988)
  143.

\bibitem{LopesCardoso:2002hd}
G.~Lopes~Cardoso {\em et.~al.}, {\it {Non-Kaehler string backgrounds and their
  five torsion classes}},  {\em Nucl. Phys.} {\bf B652} (2003) 5--34
  [\href{http://arXiv.org/abs/hep-th/0211118}{{\tt hep-th/0211118}}].

\bibitem{Aldazabal:2007sn}
G.~Aldazabal and A.~Font, {\it {A second look at N=1 supersymmetric $AdS_4$
  vacua of type IIA supergravity}},  {\em JHEP} {\bf 02} (2008) 086
  [\href{http://arXiv.org/abs/0712.1021}{{\tt 0712.1021}}].

\bibitem{Tomasiello:2007eq}
A.~Tomasiello, {\it {New string vacua from twistor spaces}},  {\em Phys. Rev.}
  {\bf D78} (2008) 046007 [\href{http://arXiv.org/abs/0712.1396}{{\tt
  0712.1396}}].

\bibitem{Koerber:2008rx}
P.~Koerber, D.~L{\"u}st and D.~Tsimpis, {\it {Type IIA AdS4 compactifications
  on cosets, interpolations and domain walls}},  {\em JHEP} {\bf 07} (2008) 017
  [\href{http://arXiv.org/abs/0804.0614}{{\tt 0804.0614}}].

\bibitem{Chatzistavrakidis:2008ii}
A.~Chatzistavrakidis, P.~Manousselis and G.~Zoupanos, {\it {Reducing the
  Heterotic Supergravity on nearly-Kahler coset spaces}},  {\em Fortschr.
  Phys.} {\bf 57} (2009) 527--534 [\href{http://arXiv.org/abs/0811.2182}{{\tt
  0811.2182}}].

\bibitem{Hull:2004in}
C.~M. Hull, {\it {A geometry for non-geometric string backgrounds}},  {\em
  JHEP} {\bf 10} (2005) 065 [\href{http://arXiv.org/abs/hep-th/0406102}{{\tt
  hep-th/0406102}}].

\bibitem{Dabholkar:2005ve}
A.~Dabholkar and C.~Hull, {\it {Generalised T-duality and non-geometric
  backgrounds}},  {\em JHEP} {\bf 05} (2006) 009
  [\href{http://arXiv.org/abs/hep-th/0512005}{{\tt hep-th/0512005}}].

\bibitem{Grana:2006kf}
M.~Grana, R.~Minasian, M.~Petrini and A.~Tomasiello, {\it {A scan for new N=1
  vacua on twisted tori}},  {\em JHEP} {\bf 05} (2007) 031
  [\href{http://arXiv.org/abs/hep-th/0609124}{{\tt hep-th/0609124}}].

\bibitem{Hull:2006qs}
C.~M. Hull, {\it {Global Aspects of T-Duality, Gauged Sigma Models and T-
  Folds}},  {\em JHEP} {\bf 10} (2007) 057
  [\href{http://arXiv.org/abs/hep-th/0604178}{{\tt hep-th/0604178}}].

\bibitem{Hull:2006va}
C.~M. Hull, {\it {Doubled geometry and T-folds}},  {\em JHEP} {\bf 07} (2007)
  080 [\href{http://arXiv.org/abs/hep-th/0605149}{{\tt hep-th/0605149}}].

\bibitem{Cvetic:2007ju}
M.~Cvetic, T.~Liu and M.~B. Schulz, {\it {Twisting K3 x T**2 orbifolds}},  {\em
  JHEP} {\bf 09} (2007) 092 [\href{http://arXiv.org/abs/hep-th/0701204}{{\tt
  hep-th/0701204}}].

\bibitem{Bergman:2007qq}
A.~Bergman and D.~Robbins, {\it {Ramond-Ramond Fields, Cohomology and
  Non-Geometric Fluxes}},  \href{http://arXiv.org/abs/0710.5158}{{\tt
  0710.5158}}.

\bibitem{Covi:2008cn}
L.~Covi {\em et.~al.}, {\it {Constraints on modular inflation in supergravity
  and string theory}},  {\em JHEP} {\bf 08} (2008) 055
  [\href{http://arXiv.org/abs/0805.3290}{{\tt 0805.3290}}].

\bibitem{Covi:2008ea}
L.~Covi {\em et.~al.}, {\it {de Sitter vacua in no-scale supergravities and
  Calabi-Yau string models}},  {\em JHEP} {\bf 06} (2008) 057
  [\href{http://arXiv.org/abs/0804.1073}{{\tt 0804.1073}}].

\bibitem{Achucarro:2008fk}
A.~Achucarro, S.~Hardeman and K.~Sousa, {\it {F-term uplifting and the
  supersymmetric integration of heavy moduli}},  {\em JHEP} {\bf 11} (2008) 003
  [\href{http://arXiv.org/abs/0809.1441}{{\tt 0809.1441}}].

\bibitem{deCarlos:2009qm}
B.~de~Carlos, A.~Guarino and J.~M. Moreno, {\it {Complete classification of
  Minkowski vacua in generalised flux models}},
  \href{http://arXiv.org/abs/0911.2876}{{\tt 0911.2876}}.

\bibitem{Gibbons:1984kp}
G.~W. Gibbons, {\it {Aspects of supergravity theories}}, . Three lectures given
  at GIFT Seminar on Theoretical Physics, San Feliu de Guixols, Spain, Jun
  4-11, 1984.

\bibitem{deWit:1986xg}
B.~de~Wit, D.~J. Smit and N.~D. Hari~Dass, {\it {Residual Supersymmetry of
  Compactified D=10 Supergravity}},  {\em Nucl. Phys.} {\bf B283} (1987) 165.

\bibitem{Maldacena:2000mw}
J.~M. Maldacena and C.~Nunez, {\it {Supergravity description of field theories
  on curved manifolds and a no go theorem}},  {\em Int. J. Mod. Phys.} {\bf
  A16} (2001) 822--855 [\href{http://arXiv.org/abs/hep-th/0007018}{{\tt
  hep-th/0007018}}].

\bibitem{Steinhardt:2008nk}
P.~J. Steinhardt and D.~Wesley, {\it {Dark Energy, Inflation and Extra
  Dimensions}},  {\em Phys. Rev.} {\bf D79} (2009) 104026
  [\href{http://arXiv.org/abs/0811.1614}{{\tt 0811.1614}}].

\bibitem{Neupane:2009ws}
I.~P. Neupane, {\it {Extra dimensions, warped compactifications and cosmic
  acceleration}},  {\em Phys. Lett.} {\bf B683} (2010) 88--95
  [\href{http://arXiv.org/abs/0903.4190}{{\tt 0903.4190}}].

\bibitem{Neupane:2009br}
I.~P. Neupane, {\it {Accelerating universe from warped extra dimensions}},
  {\em Class. Quant. Grav.} {\bf 26} (2009) 195008
  [\href{http://arXiv.org/abs/0905.2774}{{\tt 0905.2774}}].

\bibitem{Acharya:2006ne}
B.~S. Acharya, F.~Benini and R.~Valandro, {\it {Fixing moduli in exact type IIA
  flux vacua}},  {\em JHEP} {\bf 02} (2007) 018
  [\href{http://arXiv.org/abs/hep-th/0607223}{{\tt hep-th/0607223}}].

\bibitem{Dall'Agata:2004dk}
G.~Dall'Agata, {\it {On supersymmetric solutions of type IIB supergravity with
  general fluxes}},  {\em Nucl. Phys.} {\bf B695} (2004) 243--266
  [\href{http://arXiv.org/abs/hep-th/0403220}{{\tt hep-th/0403220}}].

\bibitem{Behrndt:2004mj}
K.~Behrndt and M.~Cvetic, {\it {General N = 1 Supersymmetric Fluxes in Massive
  Type IIA String Theory}},  {\em Nucl. Phys.} {\bf B708} (2005) 45--71
  [\href{http://arXiv.org/abs/hep-th/0407263}{{\tt hep-th/0407263}}].

\bibitem{Bovy:2005qq}
J.~Bovy, D.~L{\"u}st and D.~Tsimpis, {\it {N = 1,2 supersymmetric vacua of IIA
  supergravity and SU(2) structures}},  {\em JHEP} {\bf 08} (2005) 056
  [\href{http://arXiv.org/abs/hep-th/0506160}{{\tt hep-th/0506160}}].

\bibitem{Kounnas:2007dd}
C.~Kounnas, D.~L{\"u}st, P.~M. Petropoulos and D.~Tsimpis, {\it {AdS4 flux
  vacua in type II superstrings and their domain- wall solutions}},  {\em JHEP}
  {\bf 09} (2007) 051 [\href{http://arXiv.org/abs/0707.4270}{{\tt 0707.4270}}].

\bibitem{Andriot:2008va}
D.~Andriot, {\it {New supersymmetric flux vacua with intermediate SU(2)
  structure}},  {\em JHEP} {\bf 08} (2008) 096
  [\href{http://arXiv.org/abs/0804.1769}{{\tt 0804.1769}}].

\bibitem{Lust:2009mb}
D.~L{\"u}st and D.~Tsimpis, {\it {New supersymmetric AdS4 type II vacua}},
  {\em JHEP} {\bf 09} (2009) 098 [\href{http://arXiv.org/abs/0906.2561}{{\tt
  0906.2561}}].

\bibitem{Lust:2009zb}
D.~L{\"u}st and D.~Tsimpis, {\it {Classes of AdS4 type IIA/IIB
  compactifications with SU(3)xSU(3) structure}},  {\em JHEP} {\bf 04} (2009)
  111 [\href{http://arXiv.org/abs/0901.4474}{{\tt 0901.4474}}].

\bibitem{Triendl:2009ap}
H.~Triendl and J.~Louis, {\it {Type II compactifications on manifolds with
  SU(2) x SU(2) structure}},  {\em JHEP} {\bf 07} (2009) 080
  [\href{http://arXiv.org/abs/0904.2993}{{\tt 0904.2993}}].

\bibitem{Louis:2009dq}
J.~Louis, D.~Martinez-Pedrera and A.~Micu, {\it {Heterotic compactifications on
  SU(2)-structure backgrounds}},  {\em JHEP} {\bf 09} (2009) 012
  [\href{http://arXiv.org/abs/0907.3799}{{\tt 0907.3799}}].

\bibitem{Danckaert:2009hr}
T.~Danckaert and J.~Louis, {\it {Type IIA orientifold compactification on
  SU(2)-structure manifolds}},  \href{http://arXiv.org/abs/0911.5697}{{\tt
  0911.5697}}.

\bibitem{Scherk:1979zr}
J.~Scherk and J.~H. Schwarz, {\it {How to Get Masses from Extra Dimensions}},
  {\em Nucl. Phys.} {\bf B153} (1979) 61--88.

\bibitem{Hull:2005hk}
C.~M. Hull and R.~A. Reid-Edwards, {\it {Flux compactifications of string
  theory on twisted tori}},  {\em Fortsch. Phys.} {\bf 57} (2009) 862--894
  [\href{http://arXiv.org/abs/hep-th/0503114}{{\tt hep-th/0503114}}].

\bibitem{Cassani:2009ck}
D.~Cassani and A.-K. Kashani-Poor, {\it {Exploiting N=2 in consistent coset
  reductions of type IIA}},  \href{http://arXiv.org/abs/0901.4251}{{\tt
  0901.4251}}.

\bibitem{Koerber:2007hd}
P.~Koerber and D.~Tsimpis, {\it {Supersymmetric sources, integrability and
  generalized-structure compactifications}},  {\em JHEP} {\bf 08} (2007) 082
  [\href{http://arXiv.org/abs/0706.1244}{{\tt 0706.1244}}].

\bibitem{Caviezel:2008ik}
C.~Caviezel, P.~Koerber, S.~K{\"o}rs, D.~L{\"u}st, D.~Tsimpis and M.~Zagermann,
  {\it {The effective theory of type IIA AdS4 compactifications on nilmanifolds
  and cosets}},  {\em Class. Quant. Grav.} {\bf 26} (2009) 025014
  [\href{http://arXiv.org/abs/0806.3458}{{\tt 0806.3458}}].

\bibitem{Grimm:2004uq}
T.~W. Grimm and J.~Louis, {\it {The effective action of N = 1 Calabi-Yau
  orientifolds}},  {\em Nucl. Phys.} {\bf B699} (2004) 387--426
  [\href{http://arXiv.org/abs/hep-th/0403067}{{\tt hep-th/0403067}}].

\bibitem{Wecht:2007wu}
B.~Wecht, {\it {Lectures on Nongeometric Flux Compactifications}},  {\em Class.
  Quant. Grav.} {\bf 24} (2007) S773--S794
  [\href{http://arXiv.org/abs/0708.3984}{{\tt 0708.3984}}].

\bibitem{Shelton:2005cf}
J.~Shelton, W.~Taylor and B.~Wecht, {\it {Nongeometric Flux
  Compactifications}},  {\em JHEP} {\bf 10} (2005) 085
  [\href{http://arXiv.org/abs/hep-th/0508133}{{\tt hep-th/0508133}}].

\bibitem{Robbins:2007yv}
D.~Robbins and T.~Wrase, {\it {D-Terms from Generalized NS-NS Fluxes in Type
  II}},  {\em JHEP} {\bf 12} (2007) 058
  [\href{http://arXiv.org/abs/0709.2186}{{\tt 0709.2186}}].

\bibitem{Buscher:1987sk}
T.~H. Buscher, {\it {A Symmetry of the String Background Field Equations}},
  {\em Phys. Lett.} {\bf B194} (1987) 59.

\bibitem{Hassan:1999mm}
S.~F. Hassan, {\it {SO(d,d) transformations of Ramond-Ramond fields and space-
  time spinors}},  {\em Nucl. Phys.} {\bf B583} (2000) 431--453
  [\href{http://arXiv.org/abs/hep-th/9912236}{{\tt hep-th/9912236}}].

\bibitem{Micu:2007rd}
A.~Micu, E.~Palti and G.~Tasinato, {\it {Towards Minkowski Vacua in Type II
  String Compactifications}},  {\em JHEP} {\bf 03} (2007) 104
  [\href{http://arXiv.org/abs/hep-th/0701173}{{\tt hep-th/0701173}}].

\bibitem{Lust:2008zd}
D.~L{\"u}st, F.~Marchesano, L.~Martucci and D.~Tsimpis, {\it {Generalized
  non-supersymmetric flux vacua}},  {\em JHEP} {\bf 11} (2008) 021
  [\href{http://arXiv.org/abs/0807.4540}{{\tt 0807.4540}}].

\bibitem{Koerber:2010rn}
P.~Koerber and S.~K{\"o}rs, {\it {A landscape of non-supersymmetric AdS vacua
  on coset manifolds}},  \href{http://arXiv.org/abs/1001.0003}{{\tt
  1001.0003}}.

\bibitem{Villadoro:2007tb}
G.~Villadoro and F.~Zwirner, {\it {On general flux backgrounds with localized
  sources}},  {\em JHEP} {\bf 11} (2007) 082
  [\href{http://arXiv.org/abs/0710.2551}{{\tt 0710.2551}}].

\bibitem{deCarlos:2009fq}
B.~de~Carlos, A.~Guarino and J.~M. Moreno, {\it {Flux moduli stabilisation,
  Supergravity algebras and no-go theorems}},
  \href{http://arXiv.org/abs/0907.5580}{{\tt 0907.5580}}.

\bibitem{Neupane:2005nb}
I.~P. Neupane and D.~L. Wiltshire, {\it {Cosmic acceleration from M theory on
  twisted spaces}},  {\em Phys. Rev.} {\bf D72} (2005) 083509
  [\href{http://arXiv.org/abs/hep-th/0504135}{{\tt hep-th/0504135}}].

\bibitem{Font:2008vd}
A.~Font, A.~Guarino and J.~M. Moreno, {\it {Algebras and non-geometric flux
  vacua}},  {\em JHEP} {\bf 12} (2008) 050
  [\href{http://arXiv.org/abs/0809.3748}{{\tt 0809.3748}}].

\bibitem{Guarino:2008ik}
A.~Guarino and G.~J. Weatherill, {\it {Non-geometric flux vacua, S-duality and
  algebraic geometry}},  {\em JHEP} {\bf 02} (2009) 042
  [\href{http://arXiv.org/abs/0811.2190}{{\tt 0811.2190}}].

\bibitem{Gray:2008zs}
J.~Gray, Y.-H. He, A.~Ilderton and A.~Lukas, {\it {STRINGVACUA: A Mathematica
  Package for Studying Vacuum Configurations in String Phenomenology}},  {\em
  Comput. Phys. Commun.} {\bf 180} (2009) 107--119
  [\href{http://arXiv.org/abs/0801.1508}{{\tt 0801.1508}}].

\bibitem{GomezReino:2008bi}
M.~Gomez-Reino, J.~Louis and C.~A. Scrucca, {\it {No metastable de Sitter vacua
  in N=2 supergravity with only hypermultiplets}},  {\em JHEP} {\bf 02} (2009)
  003 [\href{http://arXiv.org/abs/0812.0884}{{\tt 0812.0884}}].

\bibitem{Roest:2009dq}
D.~Roest, {\it {Gaugings at angles from orientifold reductions}},
  \href{http://arXiv.org/abs/0902.0479}{{\tt 0902.0479}}.

\bibitem{Dall'Agata:2009gv}
G.~Dall'Agata, G.~Villadoro and F.~Zwirner, {\it {Type-IIA flux
  compactifications and N=4 gauged supergravities}},  {\em JHEP} {\bf 08}
  (2009) 018 [\href{http://arXiv.org/abs/0906.0370}{{\tt 0906.0370}}].

\end{thebibliography}\endgroup

\end{document}